\documentclass[twocolumn]{aastex63}

\usepackage{cancel}
\usepackage{ulem}
\usepackage{natbib}
\usepackage{array}
\usepackage{booktabs}
\defcitealias{eckert2015resolve}{E15}
\usepackage{scalerel}

\defcitealias{carr2024identification}{C24}
\let\tablenum\relax
\usepackage{siunitx}

\shorttitle{Quenching of z=0 Nuggets}
\shortauthors{Derrick S. Carr}
\graphicspath{{./}}

\usepackage[normalem]{ulem} 

\begin{document}

\title{Using Machine Learning to Estimate Near-ultraviolet Magnitudes and Probe Quenching Mechanisms of $z$=0 Nuggets in the RESOLVE and ECO Surveys}

\author{Derrick S. Carr}
\affiliation{Department of Physics \& Astronomy \\ University of North Carolina at Chapel Hill\\
Chapel Hill, NC 27599 USA}

\author{Sheila J. Kannappan}
\affiliation{Department of Physics \& Astronomy \\ University of North Carolina at Chapel Hill\\
Chapel Hill, NC 27599 USA}

\author{Zackary L. Hutchens}
\affiliation{Department of Physics \& Astronomy \\ University of North Carolina at Chapel Hill\\
Chapel Hill, NC 27599 USA}
\affiliation{Department of Physics \& Astronomy, University of North Carolina Asheville, 1 University Heights, Asheville, NC 28804, USA}

\author{Mugdha S. Polimera}
\affiliation{Department of Physics \& Astronomy \\ University of North Carolina at Chapel Hill\\
Chapel Hill, NC 27599 USA}
\affiliation{Center for Astrophysics | Harvard \& Smithsonian, 60 Garden Street, Cambridge, MA 02138, USA}

\author{Mark A. Norris}
\affiliation{Jeremiah Horrocks Institute, University of Central Lancashire, Preston, PR1 2HE, UK}

\author{Kathleen D. Eckert}
\affiliation{Department of Physics \& Astronomy \\ University of North Carolina at Chapel Hill\\
Chapel Hill, NC 27599 USA}

\author{Amanda J. Moffett}
\affiliation{Department of Physics and Astronomy, University of North Georgia, 3820 Mundy Mill Rd., Oakwood GA 30566, USA}

\begin{abstract}
We present a $z=0$ census of nuggets --- compact galaxies that form via gas-rich violent disk instability --- within the luminosity- and volume-limited REsolved Spectroscopy Of a Local VolumE (RESOLVE) and Environmental COntext (ECO) surveys. We use random forest (RF) models to predict near-ultraviolet (NUV) magnitudes for ECO galaxies that lack high-quality NUV magnitudes, thereby doubling the number of ECO galaxies with reliable extinction-corrected star formation rates (SFRs) and red/green/blue classifications based on specific SFRs (sSFRs). The resulting RF-enhanced RESOLVE+ECO nugget sample allows us to analyze rare subpopulations --- green nuggets and nuggets with active galactic nuclei (AGN) --- likely associated with quenching. Green nuggets are more similar to red nuggets than to blue nuggets in halo mass ($M_{\text{halo}}$) distribution, with both red and green nuggets being found mainly at $M_{\text{halo}} \ge 10^{11.4} M_\odot$, where permanent halo quenching is predicted. At these masses, the AGN frequency for green nuggets is higher ($\text{48.2\%}^{\text{+5.3\%}}_{\text{-5.3\%}}$) than for either blue ($\text{39.2\%}^{\text{+2.9\%}}_{\text{-2.8\%}}$) or red ($\text{29.3\%}^{\text{+3.0\%}}_{\text{-2.8\%}}$) nuggets. Between $M_{\text{halo}} = 10^{11.4}-10^{12} M_\odot$, at the onset of permanent quenching, the AGN frequency for green nuggets is nearly double the frequency for blue or red nuggets, implying AGN are associated with this transition. At $M_{\text{halo}} < 10^{11.4} M_\odot$, where temporary cyclic quenching is expected, the AGN frequency for blue nuggets ($\text{7.5\%}^{\text{+1.4\%}}_{\text{-1.2\%}}$) is lower than for either green ($\text{31.3\%}^{\text{+8.7\%}}_{\text{-7.5\%}}$) or red ($\text{18.8\%}^{\text{+11.5\%}}_{\text{-7.8\%}}$) nuggets. At all masses, nuggets with AGN have reduced sSFRs and likely also atomic gas content compared to nuggets without AGN, but the quenching is more extreme below $M_{\text{halo}} = 10^{11.4} M_\odot$.

\end{abstract}

\section{Introduction} \label{sec:intro}

Dense galaxies formed by gas-rich compaction events, a.k.a. \emph{nuggets}, represent key evolutionary phases that have helped shape the local bulged galaxy population. They were first discovered as quiescent objects, known as red nuggets, at redshift $z>1.6$ \citep{cimatti2004old,trujillo2006extremely,van2008confirmation}. Subsequently,  \citet{barro2013candels} was able to identify a high-z population of star-forming nuggets, known as blue nuggets, as their likely progenitors. Shortly after, a toy model by \citet{dekel2014wet} suggested how nuggets fit into the galaxy evolution picture. In this model, blue nuggets form via \emph{compaction} events, which are defined by gas-rich violent disk instabilities driven by wet mergers or colliding gas streams. The blue nuggets then quench into red nuggets through a combination of halo quenching and internal quenching mechanisms. After this transition, \citet{dekel2014wet} suggest that red nuggets become massive ellipticals through mostly minor mergers and satellite accretion. Building on this picture, \citet{de2016fate} studied the $z\sim0.1$ bulged galaxy population and the  $z\sim1.5$ nugget population and concluded that nuggets are primordial seeds not only for massive ellipticals but also for lenticular galaxies and classical bulged spirals (see also \citealt{gao2020local}). Thus, understanding present-day spheroids requires understanding nuggets. 

Nuggets are rare in the local universe, which makes them challenging to study robustly. Observations of the evolution of massive ($M_{*}>10^{10} M_{\odot}$) nuggets in the high-z universe show that nuggets of similar mass and density become less common over cosmic time \citep{barro2013candels,damjanov2014number,charbonnier2017abundance}. At $z=0$, \citet{saulder2015dozens} searched for massive red nuggets with low effective radii and elevated velocity dispersions and identified only 76 candidates within a parent SDSS sample of $>$$230,000$ (0.03\%). That said, dwarf nuggets may still be forming today. \citet{palumbo2020linking} performed a dedicated search for compact dwarf starburst galaxies with properties consistent with blue nuggets and found that $\sim5\%$ of all dwarf galaxies are candidate blue nuggets, roughly matching theoretical expectations from the toy model of \citet{dekel2014wet}. 

At low z, nuggets are expected to quench permanently over a range in halo mass from $M_{halo} \sim 10^{11.4} M_\odot$ to $M_{halo} \sim 10^{12.1} M_\odot$ due to shock heating of cosmic gas (e.g., \citealt{zolotov2015compaction}). Virial shocks that can significantly suppress cold gas accretion are theorized to form in the inner halo above $M_{halo} \sim 10^{11.4} M_\odot$ and to become pervasive out to the virial radius above $M_{halo} \sim 10^{12.1} M_\odot$ \citep{dekel2006galaxy,zolotov2015compaction}. These two halo masses are herein labeled the \emph{threshold} scale and \emph{bimodality} scale, respectively \citep{kannappan2013connecting}. Via the stellar mass-halo mass relation (e.g., \citealt{eckert2016resolve}), the corresponding central galaxies have stellar mass $M_{*} \sim 10^{9.6}$ and $10^{10.5} M_\odot$, respectively. We note that the cosmic web geometry that permits cold streams in higher mass halos at high z does not persist at low z \citep{dekel2006galaxy}. 


Below $M_{halo} \sim 10^{11.4} M_\odot$, in the dwarf galaxy regime, \citet{zolotov2015compaction} and \citet{tacchella2016confinement} have found that simulated nuggets cyclically quench due to temporary gas depletion and outflows, resulting in $\sim0.3$ dex oscillations on the star-forming main sequence. Using the complete volume- and luminosity-limited REsolved Spectroscopy Of a Local VolumE (RESOLVE) survey in \citet[\citetalias{carr2024identification} hereafter]{carr2024identification}, we reported preliminary $z\sim0$ evidence of cyclic quenching below the threshold scale through an analysis of star formation activity in dwarf nuggets. We also reported evidence of halo quenching above the threshold scale based on the halo mass distributions of blue and red nuggets. However, we could not confirm that the halo masses of nuggets plausibly caught in the blue-to-red transition (also known as green nuggets) closely match the halo masses of red nuggets, with only 10 green nuggets in our sample. Additionally, we would like to test whether green nuggets have intermediate atomic gas-to-stellar mass (G/S) suggestive of quenching from blue to red.

Some studies have suggested that active galactic nuclei (AGN) play a role in the blue-to-red nugget transition, although their exact role is unclear. Most simulations that focus on nuggets cannot assess AGN feedback as they do not incorporate AGN physics (e.g. \citealt{zolotov2015compaction,tacchella2016confinement,primack2018deep}); they only speculate that AGN feedback plays a role in permanently quenching nuggets. In one of the only studies to use simulations to assess the role of AGN in quenching nuggets, \citet{nogueira2019compact} compared green nuggets from hydrosimulations with green nuggets from $z\sim0.8$ observations and concluded that green nuggets quench faster than typical green valley galaxies due to being in an efficient mode of AGN feedback. 

Focusing on AGN statistics, \citet{barro2013candels} found a higher frequency of X-ray detected AGN in $z\sim2$ blue nuggets ($\sim30\%$) compared to extended blue galaxies ($\sim1\%$) where both had stellar masses $M_{*} > 10^{10} M_\odot$. The authors proposed that AGN may help remove gas from nuggets after virial shock heating shuts down cold accretion. Similarly, \citet{kocevski2017candels} found that massive ($M_{*} > 10^{10} M_\odot$) blue nuggets at $z\sim2$ are 4.7$\times$ and 7.6$\times$ times more likely to host an X-ray detected AGN than red nuggets and extended blue galaxies, respectively. They argued that AGN likely contribute to the feedback energy required to produce the short quenching timescale observed for nuggets ($\sim0.5$ Gyr; \citealt{barro2013candels}). While these studies find a role for AGN in (massive) blue and red nuggets, they do not probe the AGN frequency in green nuggets relative to red and blue nuggets, which would clarify whether AGN actually coincide with the moment of transition. The $z=0$ nugget sample of \citetalias{carr2024identification} included green nuggets, but only eight above the threshold scale, yielding a green nugget AGN frequency consistent with both the red and the blue nugget AGN frequencies within the uncertainties, despite the blue nugget AGN frequency being $\sim2\times$ higher than the red nugget AGN frequency. 



Below the threshold scale, the role of AGN in nugget quenching has received little attention.  \citet{tacchella2016confinement} found that cyclic quenching in this regime is driven by compaction cycles where first gas inflow exceeds depletion/outflows, then the reverse occurs. However, they could only track outflows from SF feedback. Despite their simulations not incorporating AGN, \citet{zolotov2015compaction} speculated that AGN feedback is likely associated with blue nuggets at all masses and may help boost internal quenching in nuggets. We studied nuggets in the cyclic quenching regime in \citetalias{carr2024identification}, but our sample had only four nuggets with AGN below the threshold scale (one blue, one green, two red), too few to probe the role of AGN in temporary quenching. Evidence of AGN feedback in non-nugget dwarf galaxies has been found in several studies (e.g. \citealt{penny2018sdss,dashyan2018agn,manzano2019agn}), so an exploration of AGN quenching in dwarf nuggets is warranted. However, with AGN detections being uncommon in dwarf galaxies ($\sim5-20\%$ of dwarfs host AGN; \citealt{polimera2022resolve,mezcua2024manga}), a larger nugget census than in \citetalias{carr2024identification} is needed to probe the dwarf nugget AGN regime. 


The present study aims to develop a larger parent sample capable of answering the above open questions. The Environmental COntext (ECO) survey, a survey $\sim$$8\times$ larger than the RESOLVE survey we used for \citetalias{carr2024identification} yet similarly designed, is a logical starting point. However, the analysis of RESOLVE nuggets was contingent on classifying nuggets as blue/green/red based on extinction-corrected star formation rates (SFRs) derived from custom-processed deep (exposure time $>1000$ seconds) Galaxy Evolution Explorer (GALEX) near-ultraviolet (NUV) imaging (\citealt{eckert2016resolve}, tabulated in \citealt{hutchens2023resolve}). Unfortunately, deep GALEX NUV imaging does not exist for roughly half of the ECO footprint, and high-quality NUV magnitudes are necessary for generating accurate SFRs (see Section \ref{subsec:justifying}). That said, nearly all of ECO does have shallower GALEX NUV photometry, typically from the All-Sky Imaging Survey (AIS, \citealt{morrissey2007calibration}). If we can use machine learning that has been calibrated using the half of ECO \emph{with} high-quality custom-processed NUV photometry to predict NUV magnitudes for the other portion of ECO \emph{without} such data, then we can use all of the ECO survey for our nugget analysis.

In this work, we train random forest (RF) regression models to fill in the missing NUV data, enabling us to create a merged RESOLVE+ECO census of 1082 nuggets within a parent sample of 10018 galaxies to probe AGN and halo quenching in nuggets. We use this enhanced data set to answer three questions:

\begin{enumerate}
    \item Are the halo mass and G/S distributions of green nuggets consistent with their being nuggets caught in the process of halo quenching? 
    \item Above the threshold scale ($M_{halo} > 10^{11.4} M_\odot$), can AGN be linked to the permanent blue-to-red nugget transition?
    \item Below the threshold scale, can AGN be linked to temporary quenching in nuggets?
\end{enumerate}

In this study, we use a standard $\Lambda$CDM cosmology with $\Omega_{m} = 0.3$, $\Omega_{\Lambda} = 0.7$ and  $H_{0} = 70$ km s$^{-1}$ Mpc$^{-1}$ except where noted otherwise. 

\begin{longtable*}{p{3.5cm} p{13.5cm}}
\caption{NUV Magnitude Glossary\label{table:glossary}}\\
\toprule
\textbf{Terminology} & \textbf{Description} \\
\midrule
\endfirsthead

\toprule
\textbf{Terminology} & \textbf{Description} \\
\midrule
\endhead

$m_{\text{band}}$ & Apparent magnitude of \emph{band}. When referring to magnitudes directly taken from the ECO/GALEX pipeline or magnitudes that are used as inputs to SED fitting, we typically use $m_{\text{band}}$. The SED fitting code applies foreground extinction corrections before fitting. \\
$M_{\text{band}}$ & Absolute magnitude of band, calculated from $m_{\text{band}}$ using luminosity distance. In this study, absolute magnitudes are k-corrected and corrected for foreground Milky Way extinction. When referring to magnitudes used as features in our random forest models, we use $M_{\text{band}}$. We may also convert $m_{\text{band}}$ either directly from the GALEX/ECO database or from SED fitting outputs to $M_{\text{band}}$ for comparison purposes. \\
custom-processed magnitudes & Magnitudes from photometry as described in \citet{eckert2015resolve,eckert2016resolve} and \citet{hutchens2023resolve}. These estimates are used as an input to the SED fitting algorithm, as described in §~\ref{subsec:stellargasmass}. \\
corrected magnitudes & Magnitudes that are internal extinction corrected in addition to the usual foreground extinction and k-corrections. These estimates are derived from our SED modeling and, for NUV, are directly used to compute corrected SFRs, as described in §~\ref{subsec:stellargasmass}. \\
shallow GALEX & NUV magnitude $M_{NUV}$ from GALEX database with the highest exposure time $<1000s$. These magnitudes are converted to $M_{NUV}$ and used as inputs to the shallow and shallow+deep random forest models. \\
deep GALEX & NUV magnitude $M_{NUV}$ from GALEX database with the highest exposure time $\ge1000s$. These magnitudes are converted to $M_{NUV}$ and used as inputs to the deep and shallow+deep random forest models. \\
GALEX pipeline magnitudes & Umbrella term for either shallow or deep GALEX $M_{NUV}$ or $M_{NUV}$ taken from the GALEX database without custom processing. \\
shallow-predicted $M_{NUV}$ & $M_{NUV}$ predicted using the shallow random forest model. \\
deep-predicted $M_{NUV}$ & $M_{NUV}$ predicted using the deep random forest model. \\
shallow+deep-predicted $M_{NUV}$ & $M_{NUV}$ predicted using the shallow+deep random forest model. \\
no-GALEX-predicted $M_{NUV}$ & $M_{NUV}$ predicted using the no-GALEX random forest model. \\
RF $M_{NUV}$ & An umbrella term for $M_{NUV}$ predicted by any of our random forest models, as described in §~\ref{sec:machinelearning}. These values come from RF trained to predict custom-processed $M_{NUV}$ and may, depending on the model, use \emph{GALEX} pipeline $M_{NUV}$ as an input feature. \\
random forest-corrected $M_{NUV}$ or $M_{NUV}$ & Similar to \emph{corrected $M_{NUV}$} but where \emph{RF} $M_{NUV}$ is converted to $M_{NUV}$ and used as an input for SED modeling in place of missing \emph{custom-processed $M_{NUV}$}. Our analysis uses these corrected magnitudes to create internal extinction-corrected SFRs for galaxies that lack corrected $M_{NUV}$ from custom-processed data. \\
SED-output $M_{NUV}$ or $M_{NUV}$ & NUV magnitudes derived from SED modeling that do not reflect internal extinction- corrections but do include foreground extinction and k-corrections. In other words, this magnitude is the SED model's refinement of the raw \emph{custom-processed} or \emph{RF} $M_{NUV}$. \\
SED-inferred $M_{NUV}$ or $M_{NUV}$ & Similar to SED-output $M_{NUV}$ or $M_{NUV}$ but where no input $M_{NUV}$ is provided to the SED fitting algorithm. \\
\bottomrule
\end{longtable*}

\section{Data and Methods} \label{sec:data_methods}

\subsection{RESOLVE and ECO} \label{subsec:resolve_eco}


The RESOLVE survey is a $\sim53,000 Mpc^{3}$ volume- and luminosity-limited census of galaxies that is highly complete down to the dwarf regime just above $M_{*} \sim 10^{9} M_{\odot}$ \citep{kannappan2008galaxy,eckert2016resolve}. RESOLVE spans two equatorial strips that create two subvolumes, RESOLVE-A and RESOLVE-B, both limited by group redshift 4500-7000 $kms^{-1}$. RESOLVE-A spans R.A. 8.75hr to 15.75hr and decl. $0$ to $5$ degrees while RESOLVE-B spans R.A. 22hr to 3hr and decl. $-1.25$ to $+1.25$ degrees. For nugget analysis, we refer to the same version of the RESOLVE survey as used by \citetalias{carr2024identification}: RESOLVE-A was cut off at the luminosity completeness limit of $M_{r} = -17.33$ matching ECO, while RESOLVE-B was cut off at $M_{r} = -17.00$ (see \citealt{eckert2015resolve}). The full volume- and luminosity-limited RESOLVE parent survey contains 1453 galaxies. For machine learning development and calibration, we only use RESOLVE-A because RESOLVE-A is a subset of ECO. RESOLVE-A contains 959 galaxies.

The ECO survey is a $>$$400,000 Mpc^{3}$ volume- and luminosity-limited survey that is similar in design to RESOLVE, except it is constructed with mostly archival data \citep{moffett2015eco,eckert2016resolve,hutchens2023resolve}. ECO naturally overlaps with the A-semester footprint of RESOLVE and adopts RESOLVE parameters for galaxies in both surveys, but it is much larger on sky (R.A. 8.7hr to 15.82hr and decl. $-1$ to $50$ degrees) and somewhat larger in group redshift range (3000-7000 $kms^{-1}$). ECO is also complete down to M$_{r} = -17.33$. The luminosity-limited parent ECO survey contains 9640 galaxies. 




\subsection{Custom Photometry} \label{subsec:customphot}

RESOLVE and ECO both use custom-processed multi-wavelength photometry from \citet{eckert2015resolve}, \citet{moffett2015eco}, and \citet{eckert2016resolve}, tabulated and updated in \citet{hutchens2023resolve}. The magnitude extrapolation estimated total galaxy magnitude by combining the flux in fixed annuli that were created using gri coadded SDSS images, with estimated flux from methods such as an outer disk fit, curve of growth \citep{munoz2009radial}, and an outer disk color correction (see \citealt{eckert2015resolve} for more details). The methods of \citet{eckert2015resolve} return brighter magnitudes and bluer colors compared to SDSS catalog magnitudes. The NUV imaging comes from the GALEX mission, so it did not require background subtraction but still underwent magnitude extrapolation via \emph{gri} annuli. Only galaxies with total NUV exposure time $>$ 1000s have custom-processed NUV photometry from \citet{eckert2015resolve,eckert2016resolve}, so we have high-quality NUV data for $\sim97\%$ of RESOLVE-A galaxies, $\sim96\%$ of RESOLVE-B galaxies, and $\sim45\%$ of ECO galaxies. While RESOLVE and ECO have very similar photometric processing methods, RESOLVE has slightly better photometry due to by-eye inspection of the data for each galaxy. More details on the custom-processed photometry pipeline can be found in \citet{eckert2015resolve}. We refer to various NUV magnitude estimates throughout this study, so we provide a glossary for each NUV term in Table \ref{table:glossary}. Conversion between $m$ and $M$ uses the equation

\begin{equation}\label{magnitude_equation}
    m = M - 5 +5log (D_{L})
\end{equation}

where the luminosity distance, $D_L$, is the line-of-sight comoving distance in megaparsecs, \texttt{loscmvgdist}, times (1+\texttt{cz}/c) based on \citet{hutchens2023resolve}.



\subsection{Structural Parameters} \label{subsec:structure}
Our approach uses seeing-deconvolved effective radii, $R_{e}$, to identify intrinsically-small objects. \citetalias{carr2024identification} used PyProFit, a profile-fitting algorithm, to obtain $R_{e}$ for RESOLVE galaxies. Roughly $90\%$ of RESOLVE galaxies have successful PyProFit models, while the other $\sim$$10\%$ have $R_{e}$ values from nondeconvolved fitting \citep{eckert2015resolve}. We use effective radii from the Dark Energy Camera Legacy Survey (DECaLS) DR9 \citep{dey2019overview} for ECO. DECaLS provides seeing-deconvolved $R_{e}$ values derived from light profile modeling via \emph{The Tractor} algorithm \citep{lang2016tractor}. \citetalias{carr2024identification} showed that DECaLS $R_{e}$ values agree well with PyProFit $R_{e}$ values for RESOLVE galaxies that have both $R_{e}$ estimates. Roughly $92\%$ of ECO galaxies have recompute $R_{e}$ from DECaLS, and in the null cases, we used $R_{e}$ values from nondeconvolved fitting from the ECO database \citep{hutchens2023resolve}. Visual inspection of galaxies with failed light profile fitting showed that they were typically either early-stage mergers or faint irregular dwarfs, both of which are difficult to fit with standard light profiles and fortunately unlikely to overlap with nuggets. In these instances, we prefer to use the photometric structural parameters of \citet{eckert2015resolve} (cataloged in \citealt{hutchens2023resolve}) rather than enforcing standard light profile models for galaxies with nonstandard profiles. We have previously shown that the nondeconvolved $R_{e}$ values agree well with the deconvolved values except at very small $R_{e}$, and that PyProFit and Tractor $R_{e}$ values are consistent (see \citetalias{carr2024identification} Section 2.5 and Figures 2 and 3). All radii are converted to units of kiloparsecs based on their Hubble distance using recessional velocity $cz$ as tabulated in \citet{hutchens2023resolve}. We also use the $\mu_{\Delta}$ parameter from \citet{kannappan2013connecting}, which acts as a quantitative morphological metric that has been shown to discern between quasi-bulgeless, bulge+disk, and bulge-dominated galaxies.

\subsection{Environment Metrics} \label{subsec:groupfind}

For this study, environment metrics come mainly from \citet{hutchens2023resolve}, which used a four-step group-finding algorithm that results in higher completeness compared to typical friends-of-friends algorithms, where completeness is defined as percentage of galaxies in the true group dark matter halo that the algorithm finds in the best-matched identified group, as measured using mock catalogs from simulations in which the true groups/dark matter halos are known. \citet{hutchens2023resolve} provided group assignments, halo abundance matching-derived halo masses, and central/satellite flags. We have also used a “flyby” flag from \citetalias{carr2024identification} that identifies flyby galaxies as galaxies inside of a halo with $M_{halo} < 10^{12} M_\odot$ and within 1.5$\times$ $R_{vir}$ of a halo with $M_{halo} \ge 10^{12} M_\odot$.

\subsection{Stellar Masses, Atomic Gas Masses, SFRs, and Active Galactic Nuclei} \label{subsec:stellargasmass}


For RESOLVE, stellar masses and extinction- and k-corrected magnitudes (including NUV) are derived from spectral energy distribution (SED) modeling (\citealt{kannappan2007systematic,kannappan2009s0,kannappan2013connecting}). The SED model grid is built using a Chabrier IMF \citep{chabrier2003galactic} and stellar population models from \citet{bruzual2003stellar}. We used a variant of the \citet{kannappan2013connecting} modeling code adopted by \citet{eckert2015resolve}, which differs from \citet{kannappan2013connecting} only in that it rejects middle-aged young populations with ages $\ge1.4$ Gyr \citep{eckert2017erratum}.

ECO also has internal and foreground extinction- and k-corrected magnitudes and stellar masses derived from the same SED modeling performed for RESOLVE. In this study, we reran the SED modeling using RF $M_{NUV}$ wherever custom-processed $M_{NUV}$ is unavailable, with RF $M_{NUV}$ coming from a sSFR machine trained on the available custom-processed NUV in \citet{hutchens2023resolve}. The RF algorithms are explained at length in § \ref{sec:machinelearning}. When rerunning the SED algorithm using RF-predicted magnitudes or GALEX pipeline magnitudes, we used a fixed NUV uncertainty of $\sim0.14$ mag to allow for random and systematic uncertainties typical of GALEX NUV photometry. We note that for galaxies that do have GALEX NUV data, the NUV uncertainty does not depend strongly or monotonically on magnitude, suggesting that a fixed uncertainty is a reasonable approximation. This new SED modeling does return new stellar mass estimates and corrected magnitudes, a few of which we tabulate in Section \ref{subsec:machinelearning} for future reference, but we only need the corrected NUV. Our analysis does employ stellar masses, but we have chosen to use the stellar masses provided in \citet{hutchens2023resolve} for consistency with all other parameters within ECO DR3. We have confirmed that stellar masses derived from SED modeling that use RF $M_{NUV}$ are consistent with stellar masses for the same subset of galaxies as tabulated in ECO DR3 \citep{hutchens2023resolve}: the median stellar mass offset between the two sets of stellar masses is 0.01 dex. Additionally, $97\%$ of the stellar masses derived from SED modeling in this work are within 0.15 dex of their stellar mass in \citet{hutchens2023resolve} (i.e., within the typical stellar mass uncertainty) and $99.5\%$ are within 0.25 dex.

Following SED modeling, SFRs were calculated for ECO and RESOLVE using both the new internal extinction-corrected NUV magnitudes and mid-IR W3+W4 magnitudes from the Wide-field Infrared Survey Explorer as described in M. S. Polimera et al. (2025, in preparation). The NUV and W3+W4 bands are used to compute a nondusty SFR (prescription from \citealt{wilkins2012accuracy}) and dusty SFR (prescription from \citealt{jarrett2012extending}), respectively. These SFRs were then combined to infer a total SFR following \citet{buat2011goods}. 

We also use HI gas masses to compute G/S ratios. Gas masses for RESOLVE were previously derived from non-flux-limited, single-pointing 21cm line data from the Arecibo and Green Bank Telescopes as well as archival flux-limited data, primarily from ALFALFA (\citealt{stark2016resolve}, updated in \citealt{hutchens2023resolve}, see \citealt{haynes2018arecibo} for ALFALFA). \citet{stark2013fueling} checked the relative flux calibrations of the ALFALFA data and the single-pointing Arecibo and GBT data and found them to be consistent (D. Stark, private communication). The non-flux-limited Arecibo and Green Bank observations were performed only for RESOLVE. For ECO galaxies outside RESOLVE-A, gas masses are based on ALFALFA-100 catalog data \citep{haynes2018arecibo} with value-added data products provided in \citet{hutchens2023resolve}. For both surveys, the best gas estimates for individual galaxies were derived from a combination of 21cm detections, deconfused detections, upper limits, and the photometric gas fraction technique, as detailed in \citet{hutchens2023resolve}. As ALFALFA is a flux-limited survey, the HI masses for ECO galaxies outside RESOLVE-A are more often derived from the photometric gas fraction technique than the HI masses for RESOLVE galaxies are.
%

This work also uses an AGN catalog for RESOLVE and ECO as developed in M. S. Polimera et al. (2025, in preparation), which extends the work of \citet{polimera2022resolve}. In short, this catalog includes AGN classification using traditional optical emission line methods e.g., BPT (Baldwin-Phillips-Terlevich, \citealt{baldwin1981classification}) and VO (Veilleux-Osterbrock, \citealt{veilleux1987spectral}), WISE photometry, and a new class of AGN (SF-AGN) that are mostly found in low-metallicity and starbursting dwarfs.



\subsection{SF Classification} \label{subsec:sfclassify}

We use the same specific star formation rate (sSFR) division as \citetalias{carr2024identification} to classify galaxies in the parent ECO survey into high, medium, and low star-formation categories based on double Gaussian fitting to identify a high star-formation locus, a low star-formation locus, and a medium star-formation region between the two loci in sSFR vs. stellar mass. For simplicity, we refer to these three categories as blue, red, and green, respectively.

\begin{figure*}[t!]
    \centering
    \includegraphics[width=180mm]{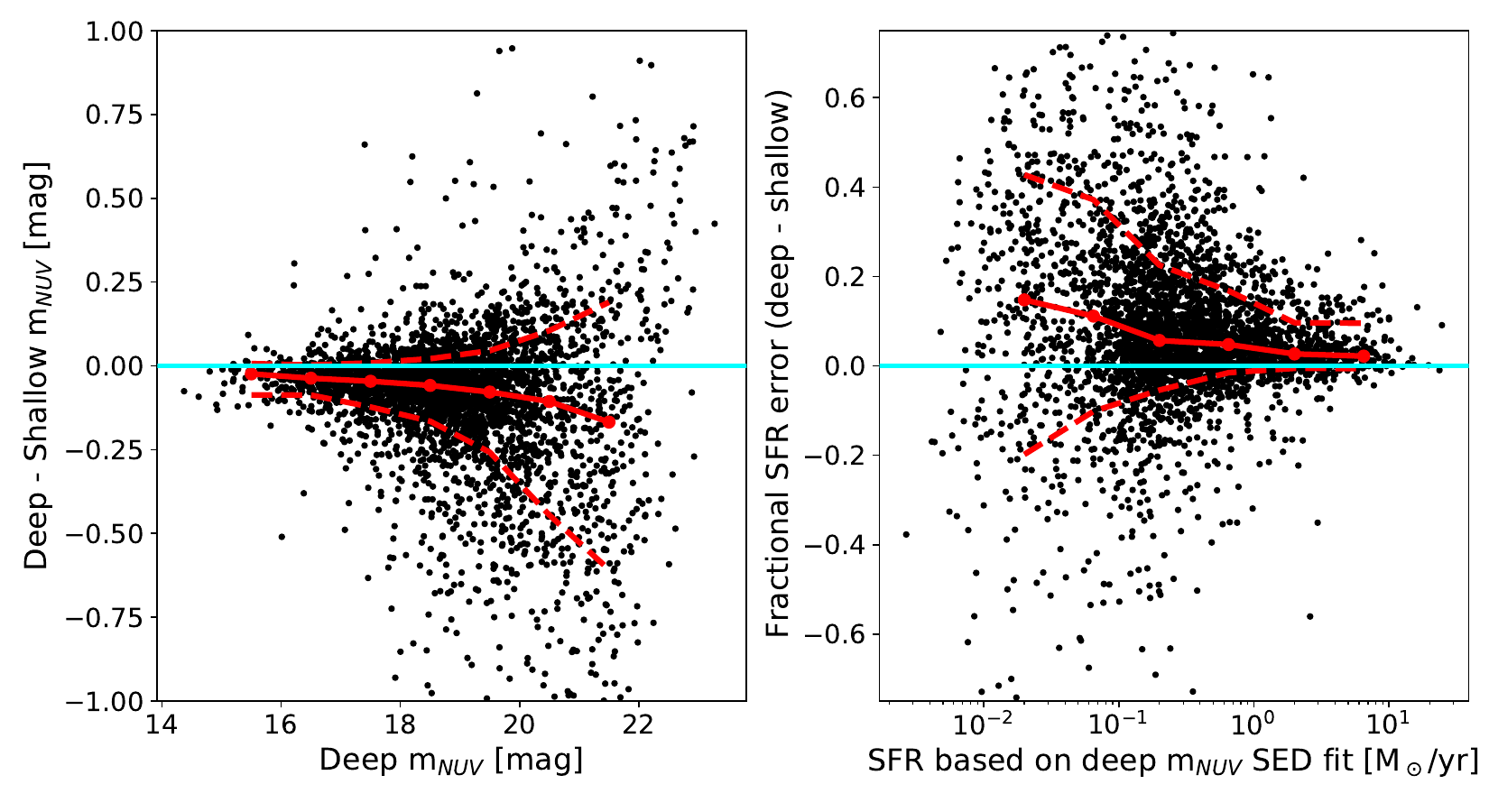}
    \caption{Comparison of pipeline GALEX data based on deep vs. shallow exposures ($>1000s$ or $<1000s$, respectively). In both plots, black dots represent ECO galaxies with both deep and shallow GALEX data, and the cyan line marks y=0. The solid red lines indicate the binned medians and the dashed red lines represent binned 84th percentiles and 16th percentiles. (Left) Deep $M_{NUV}$ - shallow $M_{NUV}$ vs deep $M_{NUV}$. Deep $M_{NUV}$ estimates are systematically brighter than shallow GALEX $M_{NUV}$ estimates, especially for fainter galaxies. (Right) Fractional error in SFRs based on SED fits vs. SFRs based on SED fits using deep GALEX data. For this panel, we separately input pipeline GALEX deep and shallow $M_{NUV}$ to the SED fitting code and use the resulting corrected $M_{NUV}$ values to calculate SFR$_{deep}$ and SFR$_{shallow}$, respectively. The y-axis fractional error is (SFR$_{deep}$ - SFR$_{shallow}$) / SFR$_{deep}$. 
    \label{fig:GALEX_comp}}
\end{figure*}

\subsection{Nugget Selection} \label{subsec:selection}

We applied the selection criteria from \citetalias{carr2024identification} to the RF-enhanced parent ECO survey and added it to our previous RESOLVE selection to create a parent RESOLVE+ECO survey. As in \citetalias{carr2024identification}, we required nugget candidates to be (1) offset to sizes below the $R_{e}$-$M_{*}$ relation for red galaxies, (2) central galaxies in their groups, and (3) nonflybys as defined in § \ref{subsec:groupfind}. We used the RESOLVE $R_{e}$-$M_{*}$ relation using $R_{e}$ from PyProFit for all galaxies in RESOLVE. For ECO galaxies not in RESOLVE, we used the $R_{e}$-$M_{*}$ relation based on ECO galaxies with $R_{e}$ from DECaLS. Not double-counting galaxies that appear in both surveys and excluding ECO galaxies that have  neither custom-processed $M_{NUV}$ nor RF $M_{NUV}$ from RF trained to predict custom-processed NUV data\footnote{We rejected 116 galaxies from our analysis that did not have custom-processed $M_{NUV}$ and for which we could not generate a successful RF} $M_{NUV}$ due to having \emph{u}, \emph{g}, \emph{r}, \emph{i}, or \emph{z} = 0 or \texttt{badrphot} $> 0$ (see \citealt{hutchens2023resolve})., the size of the parent RESOLVE+ECO sample is 10018, and the RESOLVE+ECO nugget sample is composed of 1082 nugget candidates (714 blue, 117 green, 251 red). This new nugget sample contains significantly more galaxies than the RESOLVE nugget sample from \citetalias{carr2024identification} (89 blue, 10 green, 42 red).

\begin{deluxetable*}{cl}
\tablecaption{Feature Importance\label{table:features}}
\tablehead{
\colhead{Model} & \colhead{Feature weights}
}
\startdata
Deep+shallow & Deep $M_{NUV}$ (90\%), Shallow $M_{NUV}$ (7\%), 90\% light radius (0.5\%), rest $< 0.5\%$ \\
Deep & Deep $M_{NUV}$ (98\%), 90\% light radius (0.6\%), M$_{u}$ (0.5\%), rest $< 0.5\%$ \\
Shallow & Shallow $M_{NUV}$ (96\%), M$_{u}$ (1.3\%), 90\% light radius (1\%), rest $< 0.5\%$ \\
No-GALEX & M$_{u}$ (58\%), HI gas mass (23\%), stellar mass (8\%), rest $< 5\%$ \\
\enddata
\end{deluxetable*}

\section{Methods for Extending ECO with Machine Learning} \label{sec:machinelearning}

In this section, we justify the need for RF $M_{NUV}$ from machine learning trained to predict custom-processed NUV data and describe our sSFR algorithms, which use deep/shallow GALEX and existing parameters within ECO as features to predict RF $M_{NUV}$ for galaxies that lack custom-processed $M_{NUV}$.

\subsection{Justifying the Use of Machine Learning} \label{subsec:justifying}

Custom-processed NUV magnitudes derived from deep GALEX observations are vital for robust extinction corrections via SED modeling. We need extinction corrections to estimate accurate UV-derived SFRs for classifying nuggets, as dust can heavily obscure UV light. This UV sensitivity to dust makes SED-derived extinction corrections themselves sensitive to input NUV magnitudes. Figure 19 in \citet{hutchens2023resolve} highlights the inconsistency of extinction corrections in ECO when custom-processed $M_{NUV}$ data are included as inputs to stellar population synthesis modeling versus when no NUV data are included as inputs. Blue galaxies show a $\sim0.3$ dex offset in extinction-corrected $u-r$ in the absence of custom-processed $M_{NUV}$ inputs.

The importance of including high-quality UV data to minimize bias in SED-derived quantities has also been highlighted in other studies. \citet{fan2017effects} performed UV/optical/IR SED fitting of stellar systems to probe the effect of WISE/GALEX data on SED modeling and found that excluding GALEX UV data returns systematically older ages and higher metallicities than when including GALEX data. They also concluded that GALEX UV bands are more crucial for SED fitting than other bands (e.g., WISE W1/W2) and that high-quality UV data are required for robust fitting. \citet{salim2016galex} derived sSFRs from SED models using UV/optical/IR for a low-redshift sample of $>7 \times 10^{5}$ galaxies and found that including UV data systematically drives sSFRs derived from SED fitting to higher values than when UV data are omitted from fitting, which is consistent with our findings. \citet{salim2016galex} also found that having deeper GALEX data reduces the random uncertainty on SFR, which suggests that having deeper GALEX data improves SED fitting. Thus, high-quality UV data are necessary to minimize bias in SED outputs.

Using low-quality GALEX data, such as the default shallow imaging of the GALEX AIS survey ($\sim$150$s$), does not yield robust $M_{NUV}$ for our purposes. \citet{eckert2016resolve} found that shallow GALEX imaging was inadequate for recovering NUV flux, so they only performed NUV photometry for ECO galaxies with GALEX exposure time $\ge 1000s$. To verify their conclusions, we have compiled deep and shallow GALEX pipeline magnitudes for 4277 galaxies with both. Figure \ref{fig:GALEX_comp} shows the difference in quality between deep ($>1000s$) GALEX pipeline data and shallow ($<1000s$) GALEX pipeline data. The median offset between deep GALEX $M_{NUV}$ and shallow GALEX $M_{NUV}$ is -0.06. The offset is best in the brightest bin (-0.02) and worst in the faintest bin (-0.17). We also compare SFRs calculated using internal extinction-/k-corrected $M_{NUV}$ based on SED fitting that used shallow vs. deep GALEX $M_{NUV}$ data. The median fractional SFR error is 0.05, with the median fractional SFR errors in the highest and lowest SFR bins being 0.02 and 0.15, respectively. The scatter within the 68th percentile lines extends to $\sim$0.2-0.4 dex at the faint end. Applying a simple offset correction to shallow GALEX $M_{NUV}$ data would not sufficiently correct for this discrepancy given the huge scatter. Furthermore, we aim to predict RF $M_{NUV}$ not only for ECO galaxies without custom-processed $M_{NUV}$ but also for those without shallow GALEX $M_{NUV}$. In what follows, we discuss sSFR models that yield RF $M_{NUV}$ data that can be used for robust SFRs.

\subsection{Random Forest Setup} \label{subsec:randomforestsetup}

We used machine learning algorithms that take physical parameters from the GALEX and ECO databases to predict RF $M_{NUV}$ from RF trained to predict custom-processed $M_{NUV}$ for ECO galaxies without custom-processed $M_{NUV}$. To start, we performed a cross-match between ECO DR3 and GALEX. We used a 5-arcsecond cross-match radius and extracted the $M_{NUV}$ and exposure time of all GALEX measurements for every ECO galaxy. For each galaxy, we used the deep GALEX $M_{NUV}$ with the highest exposure time $\ge 1000s$ as our deep input and the shallow GALEX $M_{NUV}$ with the highest exposure time $< 1000s$ as our shallow input, if available. 

We chose to use the random forest technique as RF algorithms can accept tabular data directly from the ECO database, utilize continuous or discrete features, and provide feature importances. We used the \textsc{RandomForestRegressor} algorithm from \textsc{scikit-learn}. We used the \textsc{squared\_error} metric to optimize our sSFR models as it was found to be stronger than other metrics in \citep{krick2020random}. We adopted a training-test split of 80/20 and the following parameters as features: M$_{\emph{u}}$, M$_{\emph{g}}$, M$_{\emph{r}}$, M$_{\emph{i}}$, M$_{\emph{z}}$, stellar mass, halo mass, HI gas mass, $90\%$ light radius, $\mu_{\Delta}$, and shallow/deep GALEX $M_{NUV}$ when available, where all non-GALEX properties are from \citet{hutchens2023resolve}\footnote{Though we do find differences in the predictive power of the deep and shallow RF models, our tests indicate that the numerical value of the exposure time has minimal feature importance ($<$0.02\%) when used as an input, so we consider it only indirectly via use of deep vs. shallow models.}. NIR magnitudes are not used as they are not uniformly available for all ECO galaxies. We chose to use absolute magnitudes to represent intrinsic, rather than distance-dependent, properties of galaxies. Stellar masses are also included to allow the sSFR to respond to the scatter in mass-to-light ratios. Halo mass is included to account for environment. We also provide the $90\%$ light radius in kpc and $\mu_{\Delta}$ as features to provide information on extent and morphology, respectively. While the HI gas mass is available for ECO galaxies, most of our sSFR models exclude it as a feature to keep $M_{NUV}$ independent of HI data and thereby maintain the ability to compare SFRs and gas data. Details on which sSFR models include/exclude HI gas data can be found in Section \ref{subsec:machinelearning} and Table \ref{table:features}.

\begin{figure*}[t!]
    \centering
    \includegraphics[width=180mm]{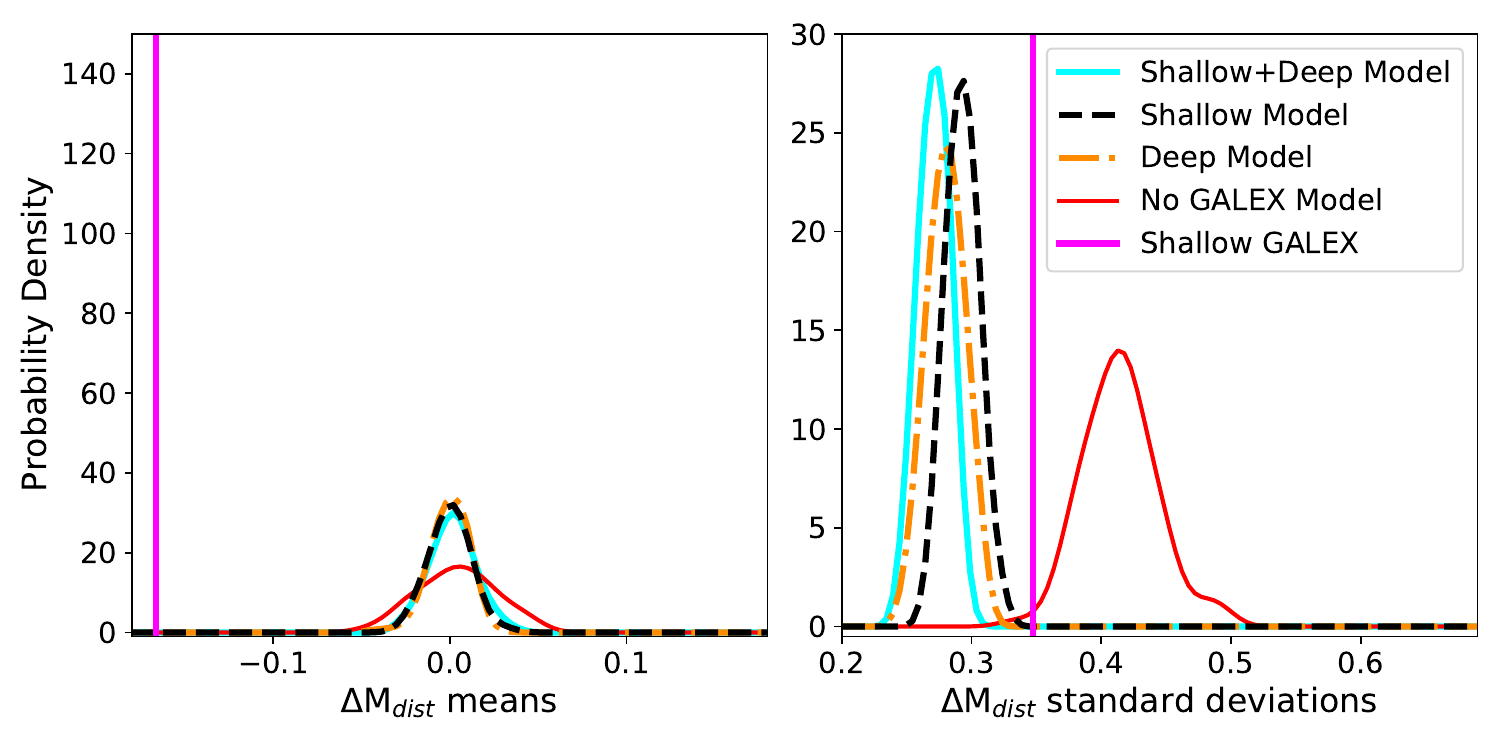}
    \caption{Performance of random forest (RF) model predictions and shallow GALEX pipeline magnitudes compared to custom-processed magnitudes. Curves show distributions of means and standard deviations of $\Delta$M$_{dist}$ (where each $\Delta$M$_{dist}$ is a distribution of the differences $\Delta$M between custom-processed $M_{NUV}$ and RF) $M_{NUV}$ for \emph{all} test galaxies in a single RF regression), where 100 independent sSFR regressions were performed for each model. Each distribution in this figure is composed of 100 data points, where a data point is the mean (left panel) or the standard deviation (right panel) of $\Delta$M$_{dist}$ from one RF regression with a unique test/train split. For comparison, vertical magenta lines mark the mean (left panel) and standard deviation (right panel) of the differences between shallow GALEX $M_{NUV}$ and custom-processed $M_{NUV}$ for ECO galaxies with both. Models that use GALEX data return comparable mean results, with slightly higher standard deviations for shallower data. The no-GALEX model has a noticeably worse performance. The shallow model corrects the large offset seen in shallow GALEX pipeline data.}
    \label{fig:100models}
\end{figure*}


Hyperparameter selection was performed in two steps. We first used a randomized search spanning a range for each of the following hyperparameters:
\begin{itemize}
    \item \textsc{n\_estimators}: {100 to 700 in steps of 10}
    \item \textsc{max\_features}: {\textsc{sqrt, auto}}
    \item \textsc{max\_depth}: {10 to 110 in steps of 10}
    \item \textsc{min\_samples\_leaf}: {2, 5, 7}
    \item \textsc{min\_samples\_split}: {1, 2, 4}
    \item \textsc{bootstrap}: {True, False}
\end{itemize}

We focused on optimizing these specific hyperparameters as they were found to be the most important for \emph{RandomForestRegressor} in \citet{mucesh2021machine}. The randomized search (\emph{sklearn.model\_selection.RandomizedSearchCV}) performed 80 cross-validated sSFR iterations using random combinations of the above hyperparameters and returned the set of hyperparameters that performed the best. From there, we performed a cross-validated grid search (\emph{sklearn.model\_selection.GidSearchCV}) centered on the set of best hyperparameters from the randomized search to determine our finalized hyperparameters. For example, if the randomized search returned \textsc{n\_estimators} = 200, the grid search would evaluate \textsc{n\_estimator} between 100 and 300 in steps of 50. The finalized hyperparameters for the sSFR algorithms were as follows: 

\begin{itemize}
    \item \textsc{n\_estimators}: 150
    \item \textsc{max\_features}: \textsc{auto}
    \item \textsc{max\_depth}: 75
    \item \textsc{min\_samples\_leaf}: 5
    \item \textsc{min\_samples\_split}: 4
    \item \textsc{bootstrap}: {True}
\end{itemize}

We found that these hyperparameter values performed well for all the sSFR models we explored.

    \subsection{Random Forest Models} \label{subsec:machinelearning}


\begin{deluxetable*}{cll}
\tablecaption{Random forest-derived $M_{NUV}$.\label{table:nuv}}
\tablehead{
\colhead{Column} & \colhead{Designation} & \colhead{Description}
}
\startdata
1 & \texttt{name} & ECO Galaxy Name \\
2 & \texttt{nuvrfshallow} & RF $M_{NUV}$ based on shallow model \\
3 & \texttt{nuvrfgalex} & RF $M_{NUV}$ based on best model, excluding no-GALEX \\
4 & \texttt{nuvrfbest} & RF $M_{NUV}$ based on best model, including no-GALEX \\
5 & \texttt{nuvrfbestflag} & Flag corresponding to best $M_{NUV}$, 0 if custom-processed $M_{NUV}$ exists \\
6 & \texttt{nuvcorr} & Corrected $M_{\text{NUV}}$ based on SED fitting where \texttt{nuvrfbest} is used as an input \\
7 & \texttt{ucorr} & Corrected $M_u$ based on SED fitting where \texttt{nuvrfbest} is used as an input \\
8 & \texttt{rcorr} & Corrected $M_r$ based on SED fitting where \texttt{nuvrfbest} is used as an input \\
\enddata
\tablecomments{
\textbf{(1)} For Column (3), best available model was used, going from best to worst: (1) shallow+deep model, (2) deep model, (3) shallow model. \textbf{(2)} Column (4) differs from Column (3) in allowing the no-GALEX model as a last choice. \textbf{(3)} For Column (5), the flag values correspond to the following: value of $-1$: could not extract a successful RF $M_{NUV}$ due to null feature(s), $0$: custom-processed $M_{NUV}$ already existed, $1$: RF $M_{NUV}$ predicted by the shallow+deep model, $2$: RF $M_{NUV}$ predicted by the deep model, $3$: RF $M_{NUV}$ predicted by the shallow model, $4$: RF $M_{NUV}$ predicted by the no-GALEX model. All RF $M_{NUV}$ come from RF trained to predict custom-processed $M_{NUV}$, not GALEX $M_{NUV}$. Absolute magnitudes may be converted to apparent magnitudes using equation~(\ref{magnitude_equation}).
}
\end{deluxetable*}

We created four sSFR models that differ based on their GALEX and HI gas data inputs. The features used for the four models are as follows:

\begin{enumerate}
    \item \emph{Shallow model}: includes M$_{\emph{u}}$, M$_{\emph{g}}$, M$_{\emph{r}}$, M$_{\emph{i}}$, M$_{\emph{z}}$, stellar mass, halo mass, $90\%$ light radius in kiloparsecs, $\mu_{\Delta}$, and shallow GALEX $M_{NUV}$
    \item \emph{Deep model}: includes M$_{\emph{u}}$, M$_{\emph{g}}$, M$_{\emph{r}}$, M$_{\emph{i}}$, M$_{\emph{z}}$, stellar mass, halo mass, $90\%$ light radius in kiloparsecs, $\mu_{\Delta}$, and deep GALEX $M_{NUV}$
    \item \emph{Shallow+deep model}: includes M$_{\emph{u}}$, M$_{\emph{g}}$, M$_{\emph{r}}$, M$_{\emph{i}}$, M$_{\emph{z}}$, stellar mass, halo mass, $90\%$ light radius in kiloparsecs, $\mu_{\Delta}$, deep GALEX $M_{NUV}$, and shallow GALEX $M_{NUV}$
    \item \emph{No-GALEX model}: includes M$_{\emph{u}}$, M$_{\emph{g}}$, M$_{\emph{r}}$, M$_{\emph{i}}$, M$_{\emph{z}}$, stellar mass, halo mass, $90\%$ light radius in kiloparsecs, $\mu_{\Delta}$, and HI gas mass
\end{enumerate}

While \citet{eckert2016resolve} derived custom-processed $M_{NUV}$ for most ECO galaxies with deep GALEX data, we found $\sim300$ galaxies with deep GALEX pipeline $M_{NUV}$ that did not have custom-processed $M_{NUV}$, hence the use of deep and deep+shallow models. For the no-GALEX model, we only use galaxies with HI gas mass derived from clean 21cm detections, strong upper limits, or deconfused observations as indicated in \citet{hutchens2023resolve}. Galaxies used to train the shallow, deep, and shallow+deep models were also used to train the no-GALEX model, assuming adequate HI gas estimates, by ignoring GALEX data. These four models were trained and tested on all galaxies that have acceptable values for every feature in their model -- the total number of train+test galaxies for the shallow+deep, deep, shallow, and no-GALEX models were 3457, 3811, 3567, and 2009, respectively. Table \ref{table:features} lists each model's top three feature importances. When available, the most important features are shallow and deep GALEX pipeline $M_{NUV}$. This result is expected as GALEX pipeline $M_{NUV}$ and RESOLVE/ECO custom-processed $M_{NUV}$ are measurements of the same intrinsic property, albeit at different levels of quality. For the No-GALEX model, the most important features were M$_{u}$, HI gas mass, and stellar mass. We note that the no-GALEX model provided very poor results when excluding HI mass, so we included HI mass.

We looked at the difference between custom-processed $M_{NUV}$ and RF $M_{NUV}$ from RF trained to predict custom-processed $M_{NUV}$ for test galaxies and found that the RF models perform well overall, with higher accuracy correlating with the availability of better GALEX data. To quantify the variations that each of our four RF models can experience, we generated 100 RF regressions using 100 unique test/train splits for each model and evaluated the typical magnitude differences in test galaxies. The magnitude difference is defined as custom-processed $M_{NUV}$ minus RF $M_{NUV}$ and is denoted as $\Delta$M, while the distribution of $\Delta$M values for \emph{all} test galaxies for a single RF run will be referred to as $\Delta$M$_{dist}$. Figure \ref{fig:100models} shows distributions composed of 100 means and 100 standard deviations, where each mean and standard deviation comes from one $\Delta$M$_{dist}$, for each of the four models. The medians of the 100 $\Delta$M$_{dist}$ means for the shallow+deep, deep, shallow, and no-GALEX models are 0.001 mag, 0.0003 mag, 0.0005 mag, and 0.002 mag, respectively\footnote{We also calculated the medians of 100 $\Delta$M$_{dist}$ \emph{medians} for the shallow+deep, deep, shallow, and no-GALEX models: they are 0.029 mag, 0.027 mag, 0.027 mag, and 0.006 mag, respectively.}. The medians of the 100 $\Delta$M$_{dist}$ standard deviations for the shallow+deep, deep, shallow, and no-GALEX models are 0.271 mag, 0.280 mag, 0.292 mag, and 0.415 mag, respectively. Thus, the shallow, deep, and shallow+deep models yield fairly similar results with increasing accuracy, while the no-GALEX model displays  increased scatter. Based on these results, we rank the performance of the models from best to worst as follows: 1) shallow+deep, 2) deep, 3) shallow, and 4) no-GALEX.  



\begin{figure*}[t!]
    \centering
    \includegraphics[width=180mm]{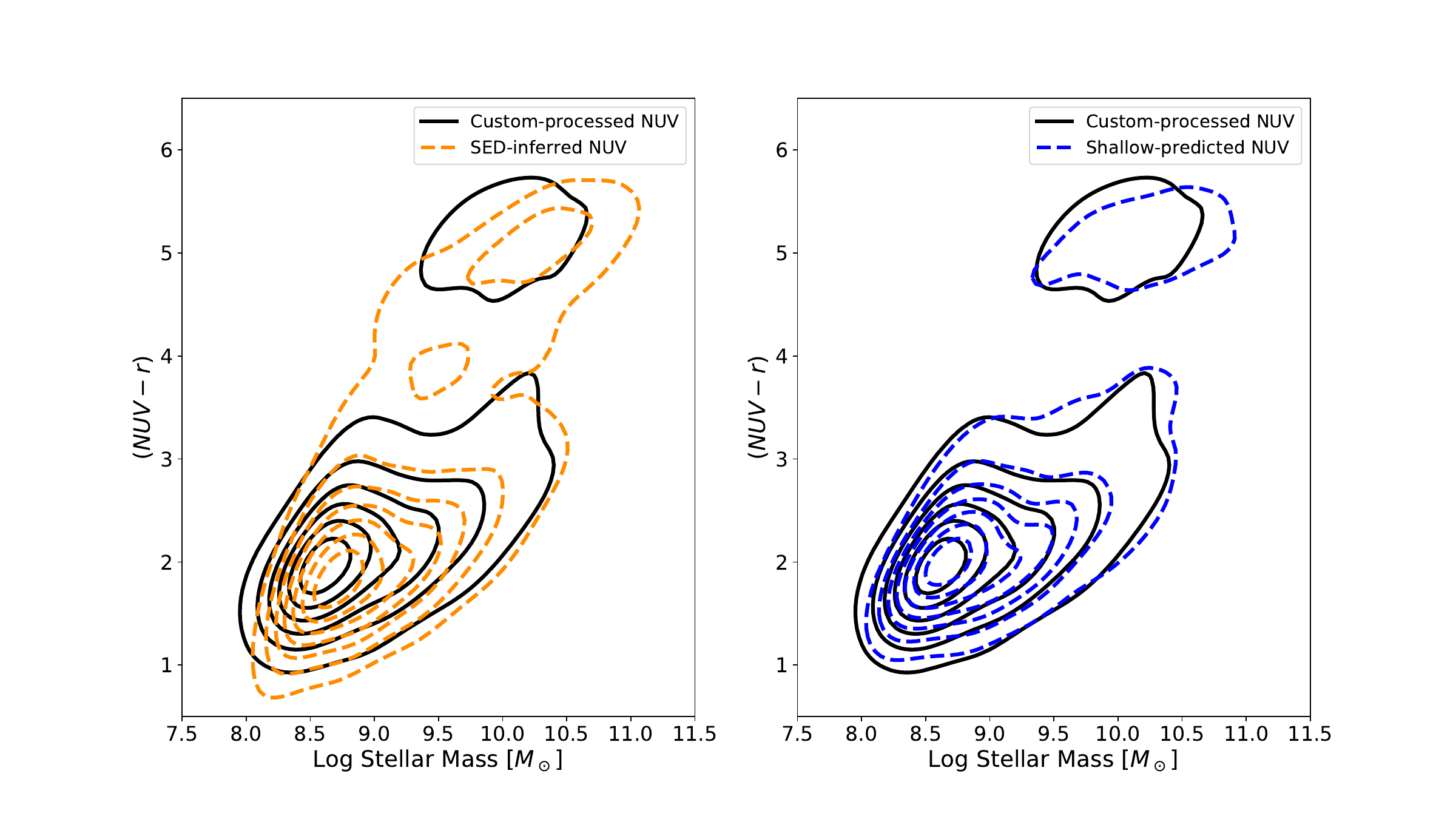}
    \caption{$NUV-r$ vs stellar mass for ECO galaxies. In both panels, the solid black contours represent ECO galaxies with custom-processed $M_{NUV}$ from \citet{eckert2016resolve} as tabulated in \citet{hutchens2023resolve}, while the dashed contours are galaxies that do not have custom-processed $M_{NUV}$ but do have shallow-predicted $M_{NUV}$. Stellar masses and m$_{r}$ are from \citet{hutchens2023resolve} in both panels. (Left) The dashed contours use $M_{NUV}$ from preexisting SED fitting for ECO \citep{eckert2016resolve}, which did not use any NUV input data for those galaxies lacking deep GALEX data. Significant offsets between the two sets of contours can be seen. (Right) The dashed contours use shallow-predicted random-forest $M_{NUV}$. The stronger agreement demonstrates that our sSFR models provide much more reliable NUV predictions than can be inferred from SED fits with no input NUV.
    \label{fig:contours}}
\end{figure*}

\begin{figure*}[t!]
    \centering
    \includegraphics[width=180mm]{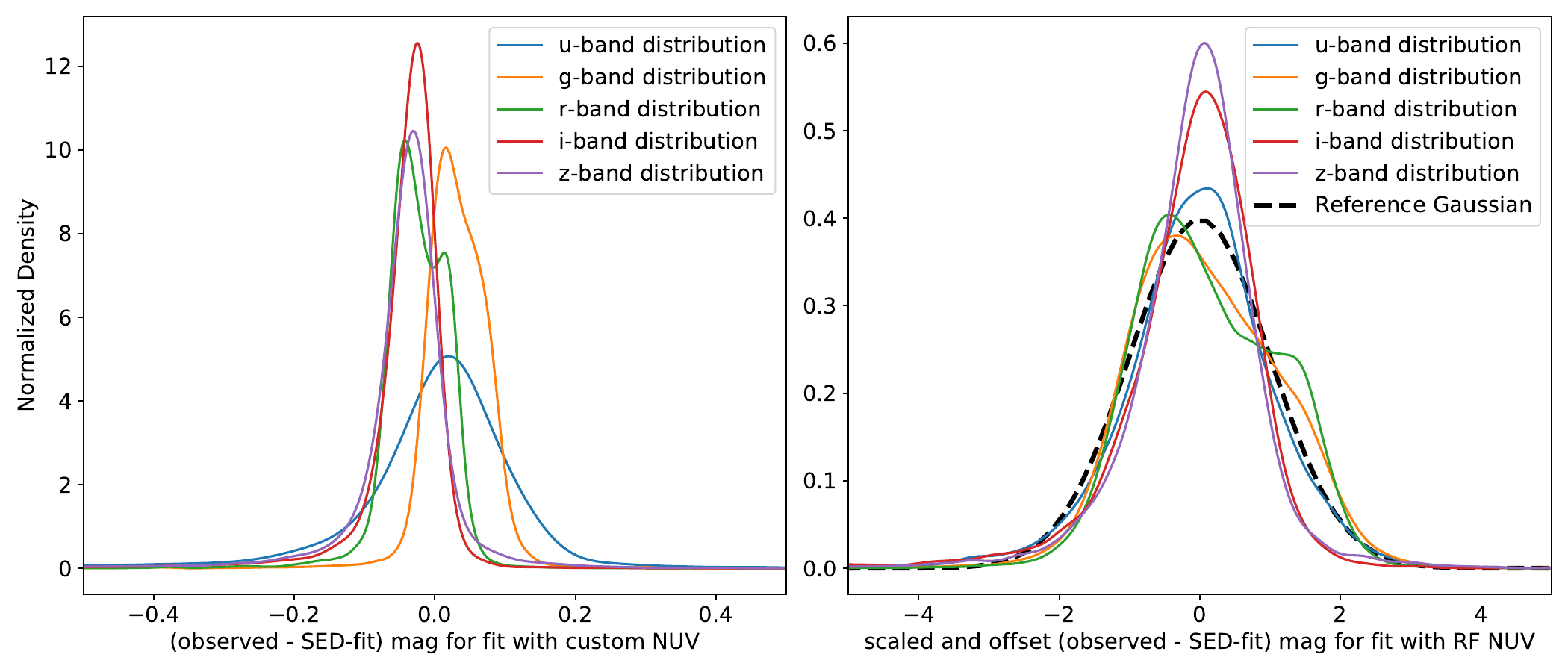}
    \caption{Comparison of \emph{ugriz} input magnitudes and output magnitudes from SED fitting for ECO galaxies. For this comparison, k-corrections and internal extinction corrections inherent in SED-output results were removed, while foreground Milky Way extinctions were matched. Left) Input \emph{ugriz} minus SED-output \emph{ugriz} magnitudes for ECO galaxies with custom-processed $M_{NUV}$. The distributions, which were created using kernel density estimation, show minor systematic $<0.1$ mag offsets (see Section \ref{subsec:machinelearning} for discussion). Right) Input \emph{ugriz} minus SED-output \emph{ugriz} magnitudes for ECO galaxies with RF $M_{NUV}$, with the systematic offsets measured in the left panel removed and the remaining residuals scaled by the inverse of the quadrature-summed systematic and observational uncertainties (see Section \ref{subsec:machinelearning} for discussion). A Gaussian with a standard deviation of 1 has been included for reference. The standard deviations of the ugriz distributions are only marginally larger than 1.0, which suggests that the quadrature summed observational and systematic input uncertainty is a reasonable proxy for the SED-output magnitude uncertainty. }
    \label{fig:banderror}
\end{figure*}

We find that our shallow-predicted $M_{NUV}$ values match custom-processed $M_{NUV}$ values better than shallow GALEX pipeline $M_{NUV}$ values do. The vertical lines in Figure \ref{fig:100models} show the mean and standard deviation of custom-processed $M_{NUV}$ minus shallow GALEX $M_{NUV}$ for ECO galaxies with both measurements. Comparing shallow GALEX pipeline $M_{NUV}$ to our shallow-predicted model $M_{NUV}$ illustrates the effectiveness of the sSFR technique. Shallow GALEX pipeline $M_{NUV}$ values have a mean offset of -0.167 mag and a standard deviation of 0.347 mag from custom-processed $M_{NUV}$ values, whereas shallow-predicted $M_{NUV}$ values have a mean offset of -0.0014 mag and a median standard deviation of 0.311 mag based on the $\Delta$M$_{dist}$ statistics in Figure \ref{fig:100models}. Not only does the shallow model almost entirely correct the flux deficit of shallow GALEX pipeline data, but it also reduces the scatter relative to custom-processed NUV data. This reduced scatter means that our shallow model performs better than would a simple offset correction based on the difference between shallow GALEX $M_{NUV}$ and custom-processed $M_{NUV}$ data.

\begin{figure*}[t!]
    \centering
    \includegraphics[width=180mm]{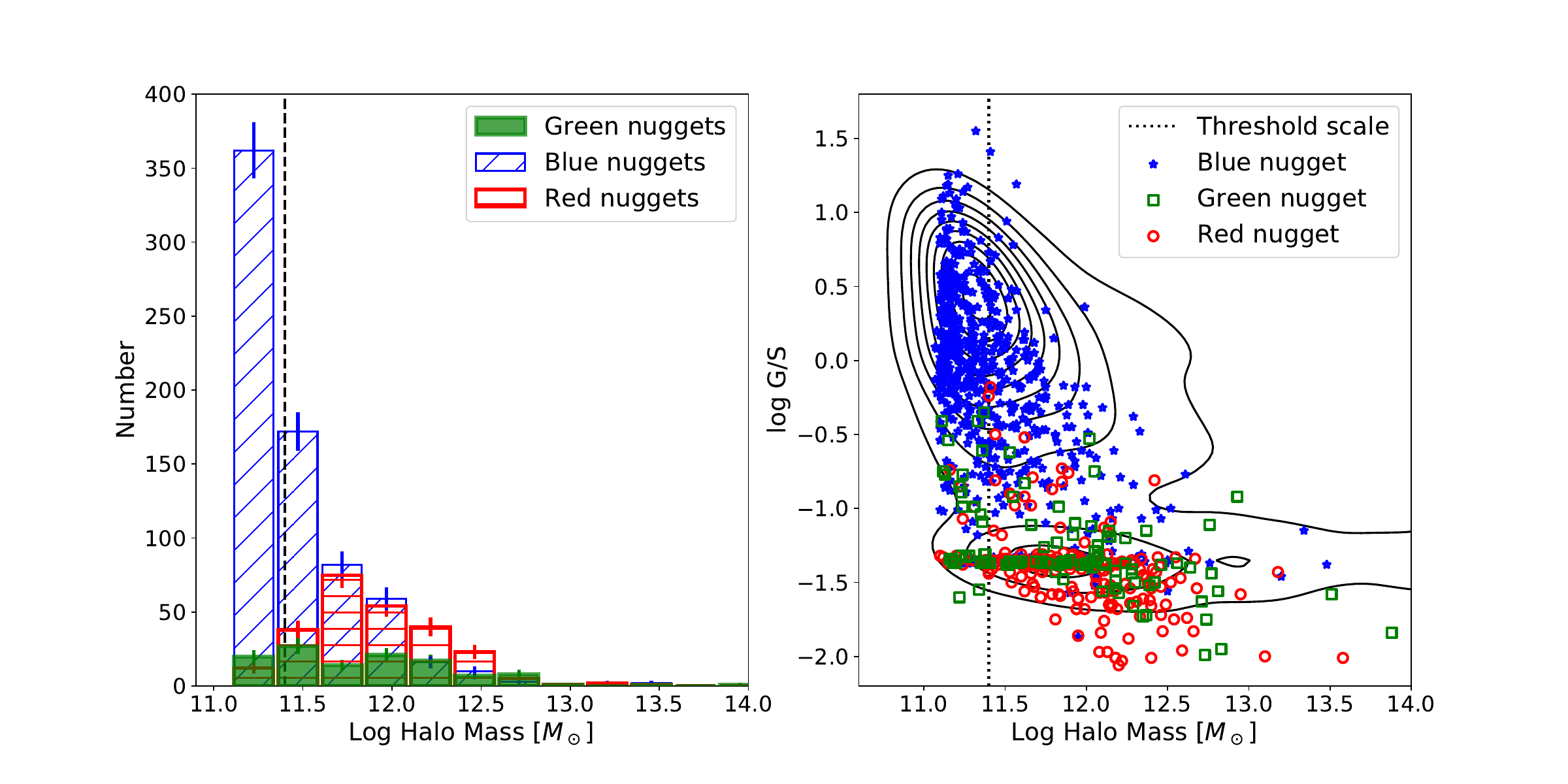}
    \caption{Halo mass distributions and G/S as a function of halo mass for ECO+RESOLVE nugget candidates. Black dashed lines in both panels represent the threshold scale. Two-sample KS tests show that green nuggets are much more similar to red nuggets than to blue nuggets in both halo mass and G/S ($Section \ref{subsec:halomassquenching}$). (Left) Error bars indicate 1$\sigma$ Poisson uncertainties. As our halo masses are derived from halo abundance matching, which assumes a monotonic relationship between group halo mass and group luminosity, the lowest mass halos are almost exclusively single-galaxy groups, and the galaxy luminosity floor described in Section \ref{subsec:resolve_eco} create an apparent halo mass floor of $M_{halo} \sim 10^{11.1} M_\odot$.} (Right) Contours represent the parent ECO+RESOLVE survey. The tight locus of green and red nuggets at log G/S $\sim 1.3$ reflects HI mass estimates based on upper limits clustered at $\sim5\%$ of the stellar mass, as described in \citet{hutchens2023resolve}.  
    \label{fig:halo}
\end{figure*}

\begin{figure}[t!]
    \centering
    \includegraphics[width=90mm]{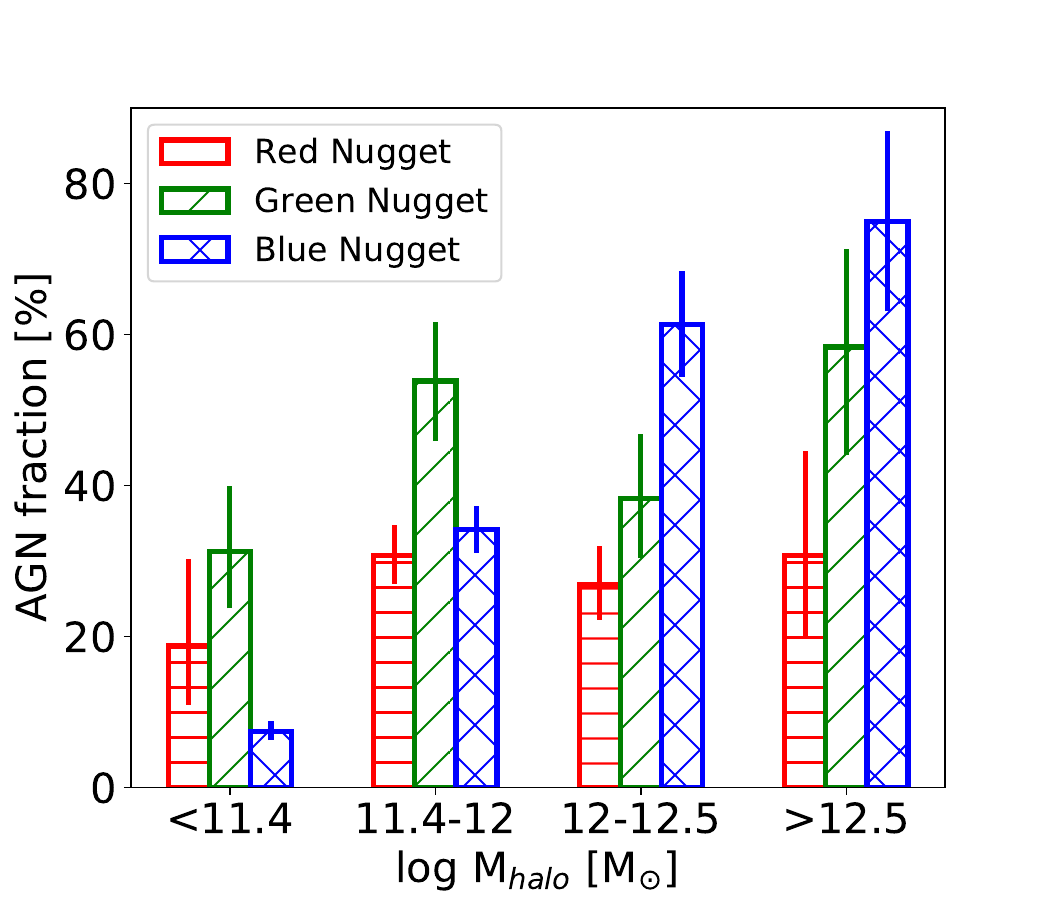}
    \caption{AGN frequencies of blue, green, and red nuggets in four halo mass bins. Binomial $1\sigma$ uncertainties are provided for each bar. In the first mass bin, blue nuggets have a notably lower AGN frequency than red or green nuggets. Green nuggets have notably higher AGN frequency than red or blue nuggets in the second mass bin, possibly due to AGN feedback accelerating quenching just above the threshold scale or, conversely, quenching allowing AGN to be more active/detectable. }
    \label{fig:agnmassbins}
\end{figure}

\begin{figure*}[t!]
    \centering
    \includegraphics[width=180mm]{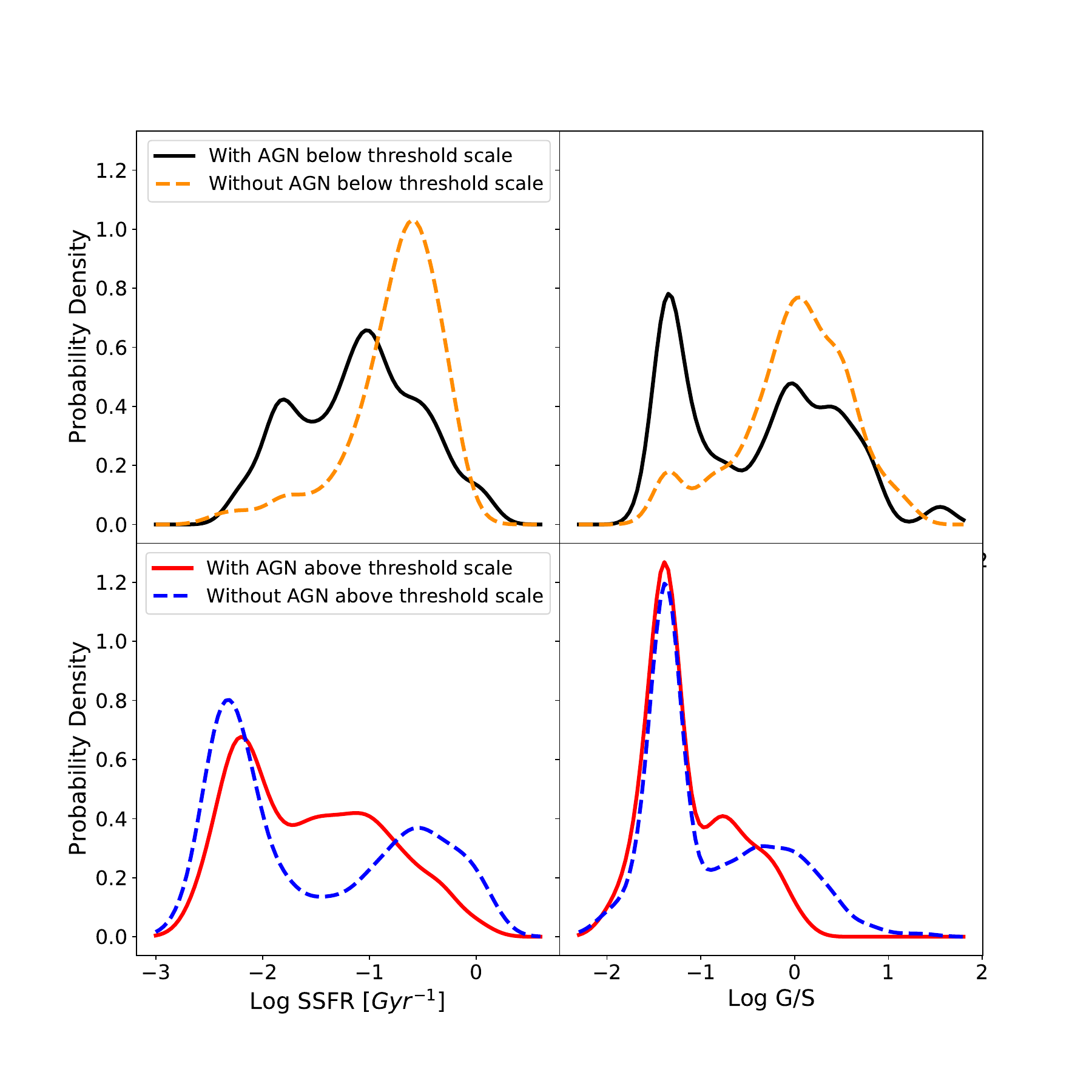}
    \caption{sSFR and G/S distributions for nuggets with AGN and nuggets without AGN in halos both above and below the threshold scale ($M_{halo} = 10^{11.4} M_\odot$). The bandwidths of the kernel density estimations are the typical sSFR uncertainty (0.20 dex) and G/S uncertainty (0.15 dex) for the left and right columns, respectively. At all masses, nuggets with AGN typically have lower G/S and sSFRs compared to nuggets without AGN, but the quenching is more pronounced below the threshold scale.
    \label{fig:dwarfagn}}
\end{figure*}

Table \ref{table:nuv} lists RF-based photometry for ECO galaxies that lack custom-processed NUV data in \citet{hutchens2023resolve}. In \texttt{nuvrfshallow}, we provide RF $M_{NUV}$ based on just the shallow model. In \texttt{nuvrfgalex}, we provide the best RF $M_{NUV}$ estimate where the order from best to worst of our models is deep+shallow, deep, and shallow. In \texttt{nuvrfbest}, we provide the same information but add the last-choice option of the no-GALEX model, which uses HI as an input feature. Finally, \texttt{nuvrfbestflag} identifies the source of the best $M_{NUV}$ estimate in \texttt{nuvrfbest}. A flag of 0 indicates custom-processed $M_{NUV}$ already existed for the galaxy, and -1 indicates that the galaxy has neither custom-processed $M_{NUV}$ nor $M_{NUV}$ from the sSFR models due to missing or unreliable input features (see footnote 1). We also provide sSFR-corrected $M_{NUV}$, M$_{u}$, and M$_{r}$ based on SED fits using \texttt{nuvrfbest} as the NUV input.

Unsurprisingly, our shallow-predicted $M_{NUV}$ values improve on the SED-inferred $M_{NUV}$ values in the current ECO database, which were inferred with no input NUV at all. Last modified in \citet{kannappan2013connecting}, the SED fitting algorithm uses available UV-optical-near infrared photometry and Bayesian stellar population modeling to infer intrinsic magnitudes and internal extinction- and k-corrections. Thus, it outputs magnitudes in bands without input magnitudes. Currently, the best ECO database estimate for $M_{NUV}$ for galaxies without custom-processed $M_{NUV}$ is this SED fit-inferred value. Figure \ref{fig:contours} shows $NUV-r$ vs stellar mass for ECO galaxies with custom-processed $M_{NUV}$ (solid black contours) compared to ECO galaxies without custom-processed $M_{NUV}$  (dashed contours in both panels). We use the same stellar masses and m$_{r}$ for all contours, so differences in the contours represent the method used to estimate $M_{NUV}$. The left panel uses SED-inferred $M_{NUV}$, and the right uses shallow-predicted sSFR $M_{NUV}$. As noted in \citet{hutchens2023resolve}, SED inference alone, with no NUV input, yields unusably inconsistent colors. In contrast, the contours derived from shallow-predicted sSFR $M_{NUV}$ follow the contours derived from custom-processed $M_{NUV}$. Thus, we have successfully enhanced ECO by providing random-forest $M_{NUV}$ estimates that are comparable in quality to custom-processed $M_{NUV}$ data.

To interpret the possible influence of cosmic variance on our machine learning models, we have separated the parent ECO survey into 10 equal-size subvolumes split by Right Ascension, then re-created the four sSFR models 10 times where each time one subvolume is excluded from the training. Using a test set composed of a random selection of 20\% of all galaxies and a training set composed of all nontest galaxies in the jackknifed volume (i.e., the full ECO volume with one subvolume removed), we found that all the models trained on different jackknifed volumes perform comparably to each other. For example, the means of the difference between the observed and predicted magnitudes for test galaxies in the jackknifed volumes for the shallow model ranged from -0.008 to -0.002. The standard deviations of the differences between the observed and predicted magnitudes for test galaxies for the shallow model ranged from 0.302 to 0.309. The No-GALEX, deep, and shallow+deep models also show similar tightness in means and standard deviations for all jackknifed volumes. Given the strong agreement in sSFR predictions between all the jackknifed volumes, we conclude that cosmic variance does not strongly influence our machine learning models. 

We fed these RF $M_{NUV}$ estimates to the SED fitting algorithm and found that the resulting uncertainties on SED-derived output magnitudes were roughly consistent with the uncertainties for the input magnitudes to SED modeling. Input photometric uncertainties were derived as described in \citet{eckert2015resolve} (and tabulated in \citealt{hutchens2023resolve}). The SED model adds an additional uncertainty component, which is an estimate of the reproducibility between different surveys due to differences in background subtraction \citep{blanton2011improved} and other methodologies such as extrapolation. These added components are 0.1 mag in NUV, 0.046 mag in \emph{u}, 0.022 mag in \emph{griz}, and 0.1 mag in JHK. Figure \ref{fig:banderror} shows the difference of input magnitude minus SED-fit output magnitude (the latter without k-corrections or internal extinction corrections to match the input magnitude) for the \emph{ugriz} bands, with separate panels for galaxies with custom-processed $M_{NUV}$ (left) vs. with RF $M_{NUV}$ (right) as the input data for SED fitting. The left panel reveals small systematic offsets for ECO galaxies with custom-processed $M_{NUV}$, where the median offsets of the \emph{ugriz} distributions are 0.072, 0.033, 0.033, 0.034, and 0.041, respectively. These values are only slightly larger than the median observational \emph{ugriz} uncertainties of 0.056, 0.024, 0.024, 0.026, and 0.034, respectively. For the right panel, we subtracted these measured systematic offsets from RF $M_{NUV}$, then scaled the remaining input-output differences by the inverse of the quadrature summed input uncertainty. The resulting ugriz distributions have standard deviations of 1.13, 1.09, 1.02, 1.04, and 0.96, respectively. This result suggests that the uncertainties associated with SED fitting increase the quadrature-summed input uncertainties by $\sim10\%$ or less and can be neglected.

Following this analysis of uncertainties, we obtained k-corrected and internal+foreground extinction-corrected RF $M_{NUV}$ values. These values were used to calculate extinction-corrected SFRs for galaxies in the parent ECO survey that lacked SFRs based on custom-processed NUV data. Finally, we generated the parent RESOLVE+ECO survey and nugget sample using all galaxies with extinction-corrected SFRs as described in § \ref{subsec:selection}.

\section{Results} \label{sec:results}

We have constructed our RESOLVE+ECO nugget sample to answer three key questions: 1) Are the halo mass and G/S distributions of green nuggets consistent with their being nuggets caught in the process of halo quenching?  2) Above the threshold scale ($M_{halo} = 10^{11.4} M_\odot$), can AGN be linked to the blue-to-red nugget transition? 3) Below the threshold scale, can AGN be linked to temporary quenching in nuggets? Below, we devote a subsection to each question.

\subsection{Halo Quenching for Green Nuggets} \label{subsec:halomassquenching}


We find that green nuggets are likely nuggets caught in the process of halo quenching. While \citetalias{carr2024identification} showed that halo mass was the driver for the blue-to-red nugget transition, \citetalias{carr2024identification} had too few green nuggets to conclusively determine whether the halo mass distribution of green nuggets is more consistent with the halo mass distribution of blue nuggets or red nuggets. With RESOLVE+ECO, we have nearly $10\times$ the number of green nuggets, allowing us to draw stronger conclusions. Figure \ref{fig:halo} (left panel) shows the halo mass distributions of blue, green, and red nuggets. The halo mass distributions of blue and red nuggets are extremely distinct, as a two-sample Kolmogorov-Smirnov (KS) test between them returns a p-value of $1.22 \times 10^{-15}$. For blue and green nuggets, a two-sample KS test returns a p-value of $7.77 \times 10^{-15}$, whereas for red nuggets and green nuggets, a two-sample KS test returns a p-value of $0.001$. Green nuggets are far more similar to red nuggets than to blue nuggets, but consistent with being a transitional population on the way to quenching. 

As shown in Figure \ref{fig:halo} (right panel), the baryonic content of our nuggets tells a somewhat consistent story. A two-sample KS test of the G/S distributions of green and red nuggets returns $p = 0.002$, whereas the same test for green and blue nuggets returns $p = 1.44 \times 10^{-15}$. When restricting our nuggets to those with HI gas mass derived from clean 21cm detections, strong upper limits, or deconfused observations, the green/red and green/blue KS tests return $p = 0.06$ and $p = 6.66 \times 10^{-16}$, respectively. Using the same gas mass restriction as above, the median log G/S for blue, green, and red nuggets are -0.01, -1.40, and -1.44, respectively, and the standard deviations of log G/S are 0.63, 0.40, and 0.34. These findings are consistent with a scenario where green nuggets, like red nuggets, have reduced gas content due to halo quenching, i.e., virial shocks suppressing cold gas refueling, explaining their suppressed sSFRs. However, green nugget G/S are not very intermediate, being roughly as low as those of red nuggets. Either their intermediate colors reflect very recent gas exhaustion, or they may have significant cold molecular gas. Compact starbursts do tend to convert HI to H$_{2}$, but also deplete and disperse H$_{2}$ via feedback, so molecular gas data would be very interesting as a window into green nugget evolution (e.g., see \citealt{stark2013fueling}).

\subsection{AGN in Nuggets Above the Threshold Scale} \label{subsec:agngiants}

Above the threshold scale ($M_{halo} \ge 10^{11.4} M_\odot$), we find that green nuggets have a higher AGN frequency than blue or red nuggets. \citetalias{carr2024identification} evaluated the AGN frequency in each of the nugget categories above the threshold scale and found AGN to be most common in blue nuggets, but the \citetalias{carr2024identification} green nugget sample consisted of only eight objects above $M_{halo} \ge 10^{11.4} M_\odot$. With the new RESOLVE+ECO nugget candidates, we performed the same analysis in the $M_{halo} \ge 10^{11.4} M_\odot$ mass regime and found that the AGN frequencies (and their associated 1$\sigma$ binomial uncertainties) in blue, green, and red nuggets are 117/298 (39.2\%$^{2.9\%}_{-2.8\%}$), 41/85 (48.2\%$^{5.4\%}_{-5.4\%}$), and 69/235 (29.4\%$^{3.1\%}_{-2.9\%}$), respectively. Thus, \citetalias{carr2024identification} missed that green nuggets have a higher overall AGN frequency than either blue or red nuggets, hinting at AGN playing a role in blue-to-red nugget quenching. 

To probe the mass dependence of AGN in nuggets, we evaluated the AGN frequencies of blue, green, and red nuggets in three halo mass bins above the threshold scale: $10^{11.4} M_\odot <= M_{halo} < 10^{12} M_\odot$, $10^{12} M_\odot <= M_{halo} < 10^{12.5} M_\odot$, and $M_{halo} >= 10^{12.5} M_\odot$. Figure \ref{fig:agnmassbins} shows the AGN frequency of nuggets in the described mass regimes. Green nuggets have a higher AGN frequency (53.8\%$^{7.8\%}_{-8.0\%}$) than blue/red nuggets (blue = 34.1\%$^{3.1\%}_{-2.9\%}$, red = 30.7\%$^{4.0\%}_{-3.8\%}$) in the bin between the threshold and bimodality scales. In the $10^{12} M_\odot <= M_{halo} < 10^{12.5} M_\odot$ bin, blue nuggets have a higher AGN frequency (61.4\%$^{7.0\%}_{-7.5\%}$) than green/red nuggets (green = 38.2\%$^{8.6\%}_{-7.9\%}$, red = 26.8\%$^{5.1\%}_{-4.6\%}$). The highest mass bin shows similar but noisier results.

We also compare the SF activity and baryonic content of nuggets with AGN to nuggets without AGN. The bottom panels of Figure \ref{fig:dwarfagn} show the sSFR and G/S distribution of nuggets in the $M_{halo} \ge 10^{11.4} M_\odot$ mass regime. The p-values from two-sample KS tests for the G/S distribution and the sSFR distribution of these two nugget subpopulations are $0.001$ and $0.0009$, respectively. When restricting our nuggets to those with HI gas mass derived from clean 21cm detections, strong upper limits, or deconfused observations, the p-values become $0.001$ and $0.001$, respectively. While nuggets without AGN above the threshold scale show bimodal distributions in both sSFR and G/S space, nuggets with AGN appear to have a somewhat higher tendency to occupy intermediate regions of sSFR and G/S, both of which are where transitional objects are likely to be found.


Overall, these results for nuggets above the threshold scale are consistent with the suggestion by  \citet{nogueira2019compact} that green nuggets are experiencing fast-mode quenching, possibly due to AGN feedback, in the presence of virial shocks that prevent cold-mode refueling. The green nugget AGN frequency may be highest right above the threshold scale because AGN feedback accelerates quenching shortly after the nugget begins experiencing virial shocks. It may also be possible that optical AGN are easier to detect in galaxies with star formation activity below the star-forming main sequence \citep{ellison2016star}. That said, blue and green nugget AGN frequencies are consistent with each other above $M_{halo} >= 10^{12.5} M_\odot$, which indicates the need for an even larger sample to improve statistics. We do find that the AGN frequency for blue nuggets is consistent with other studies of high-mass blue nuggets \citep{barro2013candels, kocevski2017candels,wang2018connecting}.

\subsection{Active Galactic Nuclei in Nuggets below the Threshold Scale} \label{subsec:agndwarfs}

Below the threshold scale, we find plausible evidence of AGN being associated with temporary quenching in some nuggets. In \citetalias{carr2024identification}, there were only four AGN-hosting nuggets out of 54 nuggets below the threshold scale, too few to assess relative AGN frequencies for blue/green/red dwarf nuggets. In the study, we find that 44 out of 464 ($9.5\%^{+1.4\%}_{-1.3\%}$) nuggets with $M_{halo} < 10^{11.4} M_\odot$ host AGN (31 blue, 10 green, three red), which is roughly consistent with the reported AGN frequency in \citetalias{carr2024identification} within much smaller uncertainties. The blue, green, and red nugget AGN fractions below the threshold scale are 31/416 (7.5\%$^{1.4\%}_{-1.2\%}$), 10/32 (31.3\%$^{8.7\%}_{-7.5\%}$), and 3/16 (18.8\%$^{11.5\%}_{-7.8\%}$), respectively. We conclude that for nuggets below the threshold scale, AGN are much more common in partially or fully quenched systems. 

We also evaluated whether differences in nugget halo mass functions may impact nugget AGN fractions. In Section \ref{subsec:halomassquenching}, it can be seen that the three nugget categories each follow distinct halo mass functions. As the blue nugget halo mass function rises toward lower mass, uncertainties in halo mass will cause more blue nuggets in low-mass halos to be scattered toward higher halo masses compared to blue nuggets in high-mass halos scattered toward lower halo masses. The scattering effect for red and green is less severe because blue nuggets have the strongest slope in their halo mass function;  so, for brevity, we describe what happens to the blue nugget population. To test the robustness of our findings to asymmetrical scatter, we added random errors to the halo masses for our final nugget sample, sampled from a Gaussian distribution with $\mu = 0$ and $\sigma = 0.3$ dex (which roughly corresponds to 1$\sigma$ uncertainties per Figure 14 of \citealt{hutchens2023resolve}). We performed this sampling 1000 times and evaluated the mean and standard deviation of the AGN fraction in each halo mass bin. After performing these Monte Carlo simulations, the mean AGN fractions of the $M_{halo} < 10^{11.4}$, $10^{11.4} M_\odot <= M_{halo} < 10^{12} M_\odot$, $10^{12} M_\odot <= M_{halo} < 10^{12.5} M_\odot$, and $M_{halo} >= 10^{12.5} M_\odot$ halo mass bins are 12.1\%, 22.4\%, 42.6\%, and 57.1\%, respectively. The standard deviations of AGN fractions for those same bins are 0.9\%, 1.8\%, 4.6\%, and 6.7\%. The trend of increasing AGN fraction with respect to halo mass seen in our results (AGN fractions of 7.5\%, 34.1\%, 61.3\%, and 75\% for the $M_{halo} < 10^{11.4}$, $10^{11.4} M_\odot <= M_{halo} < 10^{12} M_\odot$, $10^{12} M_\odot <= M_{halo} < 10^{12.5} M_\odot$, and $M_{halo} >= 10^{12.5} M_\odot$ halo mass bins, respectively) is still present, albeit weaker when adding artificial extra errors on top of the true errors to the blue nugget halo masses. Therefore, we find that the nugget halo mass function cannot produce the trend we see in nugget AGN fractions as a function of halo mass and the underlying trend is likely stronger than we measured.

The top panels of Figure \ref{fig:dwarfagn} show sSFR and G/S distributions for nuggets below the threshold scale ($M_{halo} < 10^{11.4} M_\odot$) with and without AGN. In general, dwarf nuggets with AGN have lower G/S and sSFR compared to dwarf nuggets without AGN, as seen for more massive nuggets in Section \ref{subsec:agngiants}. However, AGN/non-AGN differences are more extreme at low mass: dwarf nuggets with AGN show bimodal traits in G/S and somewhat bimodal traits in sSFR, whereas dwarf nuggets without AGN are almost entirely unquenched. Performing a two-sample KS test on the G/S distributions of nuggets with AGN and nuggets without AGN returns a p-value of 0.0005 The same test performed on the sSFR distributions of nuggets with AGN and nuggets without AGN returns a p-value of $3.37 \times 10^{-6}$. Restricting nuggets with/without AGN below the threshold scale to those with HI gas mass derived from clean 21cm detections, strong upper limits, or deconfused observations reduces the nuggets with AGN sample to just 10 nuggets, too few to effectively compare to nuggets without AGN. 

While our results link AGN to temporary quenching in nuggets in low-mass halos ($M_{halo} < 10^{11.4} M_\odot$), this result has multiple interpretations. AGN feedback may cause quenching, but our result is also consistent with the notion that the abatement of star formation feedback could allow for an AGN to turn on \citep{habouzit2017blossoms,angles2017black,bradford2018effect,polimera2022resolve}. The latter scenario is further degenerate with AGN being easier to detect in galaxies with lower star formation \citep{polimera2022resolve}. We note that only $\sim25\%$ of nuggets below the threshold scale with log sSFR $<$ -1 $Gyr^{-1}$ have detectable AGN, so it may be that other internal mechanisms, such as stellar feedback, are the primary drivers for temporary quenching. 


\subsection{Considering Cosmic Variance} \label{subsec:cosmic_variance}

We have also probed the potential impact that cosmic variance can have on our key results regarding nuggets. We used the same jackknife approach of iteratively removing subvolumes of the parent ECO+RESOLVE survey as described in Section \ref{subsec:machinelearning} to recompute our statistical results.

Concerning halo mass distributions, green nuggets still differ from blue nuggets with KS test p-values less than $1 \times 10^{-11}$, while green and red nuggets differ less, with p-values of $4 \times 10^{-4}$ to $6 \times 10^{-3}$. These results confirm that green nuggets have a halo mass distribution strongly distinct from blue nuggets and somewhat distinct from red nuggets, consistent with active quenching.

Pivoting to AGN fractions, we find that green nuggets still exhibit high AGN fractions between the threshold and bimodality scales. The green nugget AGN fraction between the threshold and bimodality scales is equal to or greater than 50\% in every jackknifed volume. Additionally, the green nugget AGN fraction between the threshold and bimodality scales is at least $\sim1.5 \times$ higher than the blue or red AGN fraction between the same scales in every jackknifed volume. Blue nuggets also still have the lowest AGN fraction below the threshold scale compared to green or red nuggets in every jackknifed volume. These results agree with our results in Section \ref{subsec:agngiants} and Section \ref{subsec:agndwarfs}.

Finally, as in Section \ref{subsec:agngiants} and Section \ref{subsec:agndwarfs}, we compared the G/S and sSFR distributions of AGN-hosting nuggets and nuggets without AGN, first for just dwarfs and then for just giants, $0.003$ and $9.69 \times 10^{-7}$ for every comparison of the G/S and sSFR distributions. This check confirms our results of AGN-hosting nuggets and non-AGN-hosting nuggets having distinct G/S and sSFR distributions.

In summary, given that our key results persist in jackknifed volumes of the parent RESOLVE+ECO survey, we conclude that our results for the full nugget census are not significantly impacted by cosmic variance.

\section{Conclusions} \label{sec:conclusions}

We have extended the nugget sample of \citetalias{carr2024identification} by using machine learning to expand the availability of high-quality NUV data for the ECO survey, thereby creating a parent RESOLVE+ECO survey $\sim8\times$ the size of the RESOLVE survey used in \citetalias{carr2024identification}. This larger $z=0$ sample of nuggets at all evolutionary phases has enabled us to better understand how nuggets quench. Specifically, we aimed to answer these three key questions: 

\begin{enumerate}
    \item Are the halo mass and G/S distributions of green nuggets consistent with their being nuggets caught in the process of halo quenching?
    \item Above the threshold scale ($M_{halo} > 10^{11.4} M_\odot$), can AGN be linked to the permanent blue-to-red nugget transition?
    \item Below the threshold scale, can AGN be linked to temporary quenching in nuggets?
\end{enumerate}

In summary: 
\begin{enumerate}
    \item We used four random forest models, each calibrated on ECO galaxies with high-quality custom-processed $M_{NUV}$, to predict high-quality random forest $M_{NUV}$ analogous to custom-processed $M_{NUV}$ for the $\sim$50\% of ECO galaxies that lack custom-processed $M_{NUV}$. The predicted $M_{NUV}$ eliminates the $\sim$0.17 mag offset and reduces the $\sim$0.35 mag scatter between shallow GALEX pipeline $M_{NUV}$ and custom-processed $M_{NUV}$. Combining the random forest and existing $M_{NUV}$ estimates in SED fitting, we have doubled the number of ECO galaxies with high-quality extinction-corrected SFRs and blue/green/red classifications based on sSFR vs. stellar mass (Section \ref{sec:machinelearning} and Figure \ref{fig:100models}).  
    
    \item Combining the random forest-enhanced ECO and RESOLVE surveys, we selected nuggets using the same selection criteria as \citetalias{carr2024identification}. In short, we required that nuggets (1) be offset to sizes smaller than the red stellar mass-effective radius relation, (2) be central galaxies, and (3) not be flyby galaxies (Section \ref{subsec:selection}).
    
    \item We find that the halo mass distribution of green nuggets is more similar to that of red nuggets, which favors halos above the threshold scale, than to that of blue nuggets, which favors low-mass halos. Green nuggets and red nuggets also have similar atomic gas content. These results suggest that green nuggets are experiencing halo quenching with suppression of cold gas accretion, though the intermediate colors of green nuggets suggest very recent or incomplete quenching. We lack data on molecular gas, which may be important in this context (Section \ref{subsec:halomassquenching} and Figure \ref{fig:halo}).
    \item Above the threshold scale ($M_{halo} \ge 10^{11.4} M_\odot$), green nuggets have an overall higher AGN frequency (48.2\%$^{5.4\%}_{-5.4\%}$) than seen for either blue (39.2\%$^{2.9\%}_{-2.8\%}$) or red nuggets (29.3\%$^{3.1\%}_{-2.9\%}$). In the narrow range between the threshold and bimodality scales, $10^{11.4} M_\odot <= M_{halo} < 10^{12} M_\odot$, green nuggets are nearly $2\times$ more likely to host an AGN than either blue or red nuggets are, hinting at the possibility that AGN play a role in permanent quenching. While AGN-hosting nuggets typically have lower G/S and sSFR than nuggets without AGN at all halo masses, we find that intermediate mildly quenched values are more common above the threshold scale (Section \ref{subsec:agngiants}, Figure \ref{fig:agnmassbins}, and Figure \ref{fig:dwarfagn}).

    
    \item AGN are less common in nuggets below the threshold scale, with the AGN frequency for blue nuggets (7.5\%$^{1.4\%}_{-1.2\%}$) lower than for green nuggets (31.3\%$^{8.7\%}_{-7.5\%}$) or red nuggets (18.8\%$^{11.5\%}_{-7.8\%}$). Below the threshold scale, AGN-hosting nuggets are associated with more extreme quenching and gas depletion than seen above the threshold scale. This result may reflect either AGN feedback or enhancement of AGN activity/detectability when star formation abates (Section \ref{subsec:agndwarfs}, Figure \ref{fig:agnmassbins}, and Figure \ref{fig:dwarfagn}).
\end{enumerate}


The expansion of ECO using machine learning allowed us to draw stronger conclusions regarding comparatively rare galaxy populations, such as green nuggets and nuggets with AGN. Using existing data in the ECO DR3 database together with GALEX pipeline $M_{NUV}$ values, our machine learning models have effectively unified the halves of ECO with and without custom-processed $M_{NUV}$. This enhancement provides uniform and robust SF data as well as multi-band extinction corrections for future studies. Thanks to machine learning, we have constructed the first dataset that is able to show that $z=0$ green nuggets are primary sites of AGN activity, correlated with both temporary quenching and permanent halo quenching. Our enhanced nugget sample showed that green nuggets have the highest AGN frequency above the threshold scale of all nuggets, especially between the threshold and bimodality scales. \citetalias{carr2024identification} missed these results due to having too few green nuggets and AGN to robustly analyze. The green nugget-AGN link may imply that AGN play a pivotal role in accelerating the blue-to-red transition, although alternative interpretations need further investigation. With nuggets being the early progenitors of nearly all bulged and early-type galaxies today \citep{de2016fate,gao2020local}, this insight into the blue-to-red nugget transition is vital for understanding the full galaxy evolution picture.

\section{Acknowledgements}

We thank Adrienne Erickcek, Carl Rodriguez, Fabian Heitsch, and Gerald Cecil for valuable feedback regarding this project. We also thank the anonymous referee for their feedback, which has made the paper significantly more compelling. This research was supported by the National Science Foundation under award AST-2007351. This work was also made possible through additional support from a UNC-CH Graduate School Summer Research Fellowship and a North Carolina Space Grant Graduate Research Fellowship.

\software{\textsc{matplotlib}, a Python library for publication quality graphics \citep{Hunter2007}; \textsc{scipy} \citep{Virtanen2020}; the \textsc{ipython} package \citep{PER-GRA:2007}; \textsc{astropy}, a community-developed core Python package for Astronomy \citep{Astropy, 2018AJ....156..123A, 2013A&A...558A..33A}; \textsc{numpy} \citep{harris2020array}; \textsc{pandas} \citep{McKinney2010, McKinney2011}}.

%

\vspace{5mm}





\bibliography{Reference}
\bibliographystyle{aasjournal}



\end{document}